\documentstyle[emulapj]{article}

\font\tenbg=cmmib10 at 10pt
\def \rvecmu{{\hbox{\tenbg\char'026}}}

\begin{document}

\title{Three-dimensional Simulations of Disk Accretion to
Inclined Dipole: I. Magnetospheric Flow at Different $\Theta$}

%\title{3D MHD Simulations of Disk Accretion to an
%  Inclined Rotator. I.
%Magnetospheric Flow at Different $\Theta$}

\author{M.M.~Romanova}
\affil{Department of Astronomy, Cornell University, Ithaca, NY
14853-6801; romanova@astro.cornell.edu}

\author{G.V.~Ustyugova}
\affil{Keldysh Institute of Applied Mathematics,
    Russian Academy of Sciences, Moscow, Russia;\\
ustyugg@spp.Keldysh.ru}

\author{A.V.~Koldoba}
\affil{Institute of Mathematical Modelling,
    Russian Academy of Sciences, Moscow, Russia;\\
    koldoba@spp.Keldysh.ru}

\author{J.V.~Wick}
\affil{Department of Astronomy, Cornell University, Ithaca, NY
14853-6801; jvw3@cornell.edu }

\author{R.V.E.~Lovelace}
\affil{Department of Astronomy, Cornell University, Ithaca, NY
14853-6801; rvl1@cornell.edu }

\medskip


\begin{abstract}

    We present results of fully
three-dimensional magnetohydrodynamic (MHD) simulations of disk
accretion to a slowly rotating magnetized star with its dipole
moment inclined at an angle $\Theta$ to the rotation axis of the
disk (which is assumed to be aligned with the spin axis of the
star).
    The main goal was to investigate the pattern of
magnetospheric flow and the disk-star interaction for a variety of
inclination angles $\Theta$. We observed that at $\Theta=0^\circ$,
the disk stops at magnetospheric radius $r_m$, and matter flows to
the star through axisymmetric funnel flows, as observed in earlier
 axisymmetric simulations. However, when the dipole moment
 of the star is
inclined, then the flow becomes non-axisymmetric.
 The non-axisymmetry
 becomes notable at very small inclination angles
  $\Theta \sim 2^\circ - 5^\circ$.  The pattern of magnetospheric
 flow is different at different $\Theta$.
     For relatively small angles,
$\Theta \lesssim 30^\circ$, the densest matter  flows to the star
mostly in two streams, which follow paths to the closest magnetic
pole.
   The streams typically co-rotate with the
star, but they may precess about the star for $\Theta \lesssim
10^\circ$.
         At intermediate angles,
$30^\circ \lesssim \Theta \lesssim 60^\circ$, the streams become
more complicated, and often split into four  streams.
        For even larger angles, $\Theta \gtrsim
60^\circ$, matter accretes in two streams, but their geometry is
different from the streams at small $\Theta$.
      Magnetic braking changes the
structure of the inner regions of the disk.
    It creates a region of lower
density (a ``gap") for  $r_m \lesssim r \lesssim 4r_m$. A ring of
 higher density forms at $r\sim r_m$ for   $\Theta \lesssim
30^\circ$. For $r \lesssim (2-3)r_m$, the azimuthal velocities are
sub-Keplerian.
    The inner region of the disk
at  $r\sim r_m$ is warped.
     The warping is due to the
tendency of matter to co-rotate with inclined magnetosphere.
  The normal of the inner
  warped part of the disk is close
to the magnetic axis of the dipole. The accreting matter brings
positive angular momentum to the (slowly rotating) star tending to
spin it up. The corresponding torque $N_z$ depends only weakly on
$\Theta$.
    The angular momentum flux to the
star near the star's surface is transported predominantly by the
magnetic field;  the matter component contributes $\sim 1\%$ of
the total flux.
   The torques $N_x$ and $N_y$ are also calculated and these may
   give a slow precession of the symmetry axis of the star.
     The angle $\Theta$ was fixed in
simulations because the time scale of its evolution is much longer
that that of the simulations.
 Results of simulations are important for understanding
the nature of classical T Tauri stars, cataclysmic variables, and
X-ray pulsars. These stars often show complicated spectral and
photometric variability patterns, which may be connected with the
structure of magnetospheric flows.
    The magnetospheric structure of stars
with different $\Theta$ can give different variability patterns in
observed light curves.
    This can provide information
about inclination angles $\Theta$ in different stars.
     A notable result of the present
simulations is the formation of multiple streams in the accretion
flows near the star for intermediate inclination angles. This may
give short-scale quasi-periodic variability in the light curves of
some stars.

\end{abstract}

\section{Introduction}

\noindent{\it 1.1.~ Observations.}~ Disk accretion to a rotating
star with a dipole magnetic field is important in a number of
objects, including classical T Tauri stars, cataclysmic variables
and X-ray binaries.
       Commonly, the
magnetic axis of the dipole is not aligned with the rotation axis
of the disk.
        The disk accretion  to
rotating neutron stars with misaligned dipole fields gives rise to
X-ray pulsars in binary systems (see, e.g., Tr\"umper et al. 1985;
Sheffer et al. 1992;  Bildsten et al. 1997; Deeter et al. 1998).
      The periodic variability in some
cataclysmic variables is thought to be due to the rotation of a
star with a misaligned dipole field
    (e.g., Livio \& Pringle 1992; Warner
1995; Warner 2000;  Wickramasinghe, Wu, \&  Ferrario 1991).
      A similar cause may give the
photometric and spectral variability of some classical T Tauri
stars (e.g., Herbst et al. 1986; Bouvier \& Bertout 1989; Johns \&
Basri 1995; Hartmann 1998; Bouvier,  et al. 1999, 2003; Petrov, et
al. 2001).

       Disk accretion to an inclined dipole rotator
occurs in different systems.
       For specificity we adopt a scaling
of the parameters appropriate for classical T Tauri Stars (CTTSs).
       Application to other objects such as
neutron stars may be done by rescaling the parameters.

         Following the discovery of  variability of
         classical T Tauri stars (CTTSs)
(Herbst et al. 1986; Bertout et al. 1988), it was suggested that
this variability may be due to rotational modulation of emission
of ``hot spots" on the surface of the star which are connected
with magnetospheric funnel flows (Camenzind 1990;  K\"onigl 1991).
           Many observations of CTTSs support
this model.
        In some cases (e.g., AA Tau), the
variability may  be due to the warp of the inner part of the disk
(Bouvier et al. 1999, 2003).
     Photometric and spectral variability
of CTTSs often shows complicated patterns which have not been
explained.
       In order to understand the nature
of the variability of different CTTSs, it is important to
calculate the detailed MHD flow pattern in
    the magnetosphere outside of
the inclined rotator.
      The magnetic field of CTTSs may not
be a pure dipole (e.g., Safier 1998, Smirnov et al. 2003).
     However,
this paper considers the case where the star's field is a dipole
field.

\begin{figure*}[t]
\epsscale{1.2} \plotone{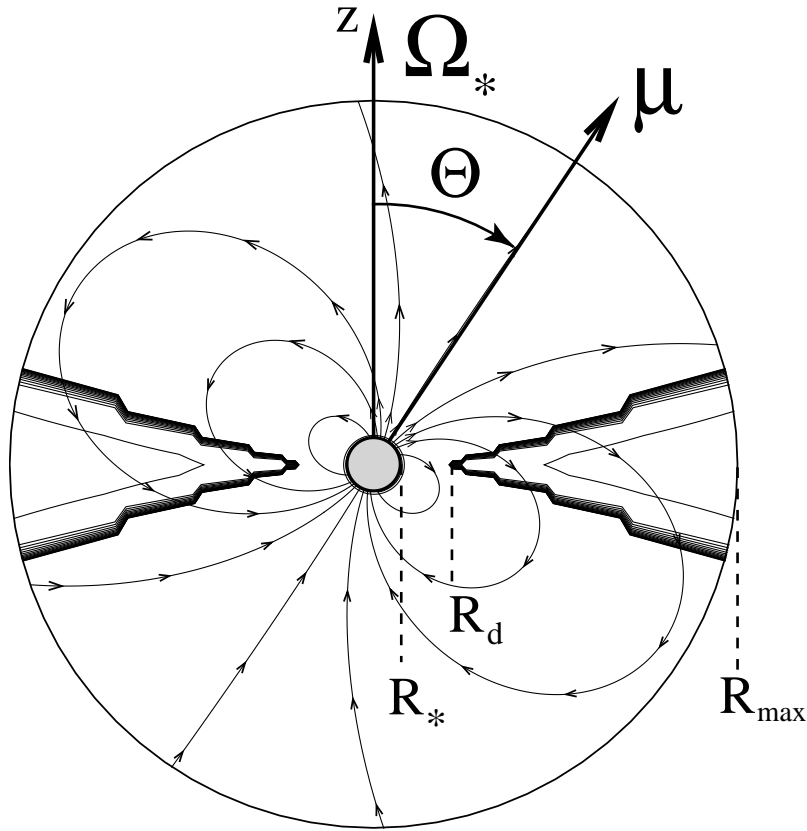} \caption{Initial conditions for
simulations. A star of radius $R_*$ is threaded with the dipole
magnetic field with the magnetic momentum ${\bf \mu}$ inclined
relative to the $z-$ axis at an angle $\Theta$. $R_{\rm max}$ - is
the size of the simulations region. The solid lines with arrows
represent magnetic field lines. The level lines of the disk are
shown. The inner radius of the disk is located at $R_d$.}
\label{Figure 1}
\end{figure*}

\smallskip

\noindent{\it 1.2.~ Theory of Disk Accretion to an Inclined
Rotator. }~ The problem of disk accretion to an inclined dipole
rotator has been investigated theoretically by a number of
authors. These works considered the the possible warping of the
inner part of the disk.

 It was noticed in 1970s that in case of diamagnetic disk
 (when magnetic field does not penetrate  the disk)
 the magnetic pressure will be
different above and below the disk so that there is a net force
perpendicular to the disk.
     This force causes  warping of the disk (Scharlemann
1978; Aly 1980).
    This early work considered the case in which the
rotational axis of the distant disk ${\bf \Omega}_d$ and spin of
the star ${\bf \Omega}_*$ coincide.
    Later, the analysis was extended to the case in which the
${\bf \Omega}_d$ and ${\bf \Omega}_*$ are in different directions
(Lipunov \& Shakura 1980; Lipunov, Semenov \& Shakura 1981; Horn
\& Kundt 1989; Lai 1999).
       The net force acting on the inner regions of the
disk (averaged over the rotation of the star) acts so as to make
the normal to the warped disk
  perpendicular to both the magnetic axis of the dipole and spin of
  the star.
      The disk is expected to precess around the spin axis
  of the star.
     Subsequently, Lai (1999) extended
this analysis to cases of imperfectly conducting disks.
  Later, it was shown that {\it viscosity} may decrease warping
  % of the disk was included
(Horn \& Kundt 1989; Lai 1999).
     Agapitou, Papaloizou  \& Terquem (1997)
     and Terquem \& Papaloizou (2000)
     conclude that bending waves
     may also lead to warping, and the amplitude of the warp
     will depend on viscosity parameter $\alpha$.

     The theoretical models typically involve  a number of
simplifying approximations. Predictions of these models should be
checked numerically. This paper, however, does not focus on this
topic, and  no special simulations were arranged to enhance the
warping.

No theories were done on the main topic of this paper - analysis
of magnetospheric flows in the inclined magnetosphere. This
problem is principally three-dimensional and needs full 3D
simulations. Some of the important problems, however, were
investigated analytically and numerically in an axisymmetric
approach, which is valuable for subsequent 3D analysis.

\smallskip{\it 2.3.~ 2D Simulations of the Disk-Star
Interaction}.~
        Different aspects of the
disk-star interaction for the case of an aligned rotator were
investigated  with two-dimenionsal, axisymmtric MHD simulations
(Hayashi, Shibata \& Matsumoto 1996; Miller \& Stone 1997; Hirose
et al. 1997).
      Special attention
was given to the problem of opening of the coronal magnetic field
lines and associated outflows (Goodson, B\"ohm \& Winglee 1999;
Fend \& Elstner 2000).
       This process was earlier investigated
theoretically by  Lovelace, Romanova, \& Bisnovatyi-Kogan (1995),
and Bardou \& Heyvaerts (1996).
      Much less attention has been
given to numerical investigation of the dipole interaction with
the inner regions of the disk and modelling of magnetospheric
flows, which were investigated analytically by Ghosh \& Lamb
(1979a,b), Shu et al. (1994), Ostriker \& Shu (1995), Wang (1995),
Li \& Wickramasinghe (1997), Koldoba et al. (2002a), and others.
      This problem was not
fully investigated by MHD simulations because of the difficulty of
creating consistent ``quiescent initial conditions.''
      In most simulations the
initial magnetic braking of the disk by the magnetospheric field
leads to the fast accretion of disk matter to the star with
velocity close to free-fall (e.g., Hayashi et al. 1996; Miller \&
Stone 1997).
      Furthermore, the strong initial twist
forms at the boundary between the disk and corona, and this acts
to initiate outflows.

      Romanova et al. (2002, hereafter R02)
developed consistent quiescent initial conditions, which lead to
slow, viscous time-scale accretion avoiding the free-fall
accretion of earlier studies.
      This opened up the
possibility of investigating the physics of the disk-star
interaction and magnetospheric flows in great detail on the
viscous time-scale.
      Many aspects of the earlier
predicted theories  were reconsidered (R02).
       It was shown that
     (1) the magnetic field
lines in the inner regions of the disk have a tendency to be
closed, while external lines may be open (see also recent paper by
K\"uker, Henning,  R\"udiger 2003);
     (2) the inner regions of the disk
experience magnetic braking and their angular velocity is
smaller/larger than Keplerian (depending on angular velocity of
the star);
     (3) the magnetic field is dominant
in the spinning-up/spinning-down of the star; (4) the main force
lifting matter up to the funnel flow is the pressure gradient
force, while magnetic force is negligibly small.
      Different aspects
of earlier proposed models were confirmed or rejected.
     With detailed axisymmetric analysis
in hand, it was easier to under take the three-dimensional MHD
simulations of the disk accretion to the inclined dipole.

     In order to understand accretion
to a misaligned rotator,
    we created a special three-dimensional MHD
simulation code based on a ``cubed sphere" grid.
       This grid has a number of advantages over
spherical or  Cartesian grids (Koldoba et al. 2002b, hereafter
K02).
     A similar grid was developed
for the surface of the sphere for geophysical applications (see,
e.g.,  Ronchi, Iacono, \& Paolucci 1996).
       In contrast with
these authors, we perform simulations in the three-space and use
as a base the Godunov-type numerical scheme which is similar to
one described by Powell et al. (1999).
      We performed
sample simulations of  three-dimensional accretion to an inclined
rotator for an inclination angle $\Theta=30^\circ$ and obtained
the first interesting results.
       We observed that after one rotation of the inner
radius of the disk, the initial axisymmetry of the disk
    is destroyed, and matter flows
to the star along two streams.

      In this paper we
present the results for the full range of inclination angles
$\Theta$.  The main goals of the simulations are the following:
(1)~ to investigate the structure of the magnetospheric flows at
different inclination angles $\Theta$, which is important for
subsequent analysis of spectral and photometric variability of
CTTSs and other stars; (2)~ to investigate the properties of the
disk surrounding the magnetosphere; (3)~ to investigate the rate
of spinning up/down of the star and its dependence on $\Theta$;
and (4)~ to investigate how inclination angles $\Theta$ may change
with time.  In future papers we will show detailed analysis of the
3D streams and hot spots,  and will derive the light curves from
simulations (Romanova et al. 2004a).

      In \S 2, the numerical model is described.
In \S 3, test simulations are presented.
      In \S 4, the main
numerical results for different $\Theta$ are discussed.
      In \S 5,
the magnetic braking of the disk is analyzed.
       In \S 6, the fluxes of matter and
    torque   to the star are calculated. In \S 7,
    the main conclusions of this work are presented and
    the possible observational consequences are discussed.

\section{Numerical Model}

       We investigate the
case in which the angular velocity of the star ${\bf \Omega=\bf
\Omega_*}$ and that of the disk are aligned and in the $\hat{\bf
z}$ direction.
       The magnetic
moment $\rvecmu$ is inclined relative to $\hat z$ axis by an angle
$\Theta$ as shown in Figure 1. We fix inclination angle $\Theta$
and do not change it during simulations, thus reflecting the fact
that the angular momentum of the star is much larger than the  angular
momentum of accreting matter.
       In general, the angular momenta of the
star and that of the disk may be misaligned, but in the present
paper we consider only the aligned case. The initial magnetic
field is considered to be a dipole field,  $ {\bf B}=3(\rvecmu
\cdot {\bf R}){\bf R}/{R^5} - \rvecmu/{R^3}$.
      We use  a reference frame $(X,Y,Z)$ rotating with
the star with the $Z$ axis aligned with the star's rotation axis.
      This rotating frame is orientated such that
$\rvecmu$ is in the $(X,Z)$ plane.

We solve the  system of ideal MHD equations.
   We decompose the
magnetic field ${\bf B}$ into the
``main" dipole component of the
star ${\bf B}_0$, and the component,
${\bf B}_1$, induced by
currents in the disk and in the corona.
   This helps to minimize
errors in calculating the magnetic
force (e.g., Tanaka 1994).  In
the case of the inclined  rotating dipole
    the dipole moment changes with time. It
rotates with angular velocity ${\bf \Omega}$ so that the ``main"
field ${\bf B}_0$ also changes with time.
    Consequently in
the induction equation there is a large
term involving ${\bf
B}_0$.
    To overcome this difficulty we use a coordinate system
rotating with angular velocity ${\bf \Omega}$, in which the
magnetic moment of the star ${\rvecmu}$ and the ``main" field
${\bf B}_0$ do not depend on time.
    The system of MHD
equations in the coordinate system
rotating with the star is
\begin{equation}
{\frac{\partial \rho}{\partial t}} + {\bf {\nabla}}\cdot (\rho{\bf
v}) = 0~,
\end{equation}
\begin{equation}
\frac{\partial (\rho{\bf  v})}{\partial t} + {\bf {\nabla}}\cdot
{T} = \rho {\bf g} + 2\rho ~{\bf v}\times{\bf \Omega} - \rho~ {\bf
\Omega}\times ({\bf\Omega}\times{\bf R})~,
\end{equation}
\begin{equation}
\frac{\partial (\rho s)}{\partial t} + {\bf {\nabla}}\cdot (\rho s
{\bf v}) = 0~,
\end{equation}
\begin{equation}
\frac{\partial {\bf B}}{\partial t} = {\bf \nabla \times} ({\bf
v}\times{\bf B})~,
\end{equation}
(K02) where $\rho$  is the plasma density, $s$ is the entropy,
$T_{ik}$ is the stress tensor with
  $T_{ik} = p\delta_{ik} +\rho v_i v_k+
({{\bf B}^2}\delta_{ik}/2-B_iB_k)/4\pi$.
   Here ${\bf v}$  is the velocity of plasma
in the rotating frame. The transformation to the  inertial
(observer's) frame is $ {\bf u} = {\bf v} + {\bf \Omega}\times{\bf
R}$.
   One can see that equation (2),
which is the Euler equation in
the rotating frame, has
new terms compared to the
usual Euler equation, namely, the Coriolis
and centrifugal forces.
    The Godunov-type numerical code was used
in the ``cubed sphere" coordinate system.
      The code was tested and verified to
give the  correct solution for
Bondi accretion to a gravitating
object with a monopole
magnetic field (K02).

\subsection{Reference Values and Example  for
T Tauri Stars}

     The reference value for distance is
denoted $R_0$, and it is taken to be the initial value of the
inner radius of the disk.
      Thus the dimensionless inner
disk radius is $R_d=1$ initially.
     We take the dimensionless radius of the star to be
$R_*=0.35$.
         The reference value for the velocity is
taken to be the Keplerian velocity at $R_0$, $v_0 =
(GM/R_0)^{1/2}$.
        The reference angular rotation
rate is $\omega_0 = v_0/R_0$, and the corresponding time-scale is
$t_0=R_0/v_0$.
          We also use the rotation period  at $r=R_0$,
$P_0=2\pi t_0 $.
         For a given magnetic field strength
at $R_0$ we can define a reference density as $\rho_0 =
B_0^2/v_0^2$ and reference pressure $p_0=\rho v_0^2$.
            A reference magnetic moment of
the star is then $\mu_0 =B_0 R_0^3$.
      Thus, the calculated
variables are: $R'=R/R_0$, $t'=t/t_0$, $\rho'=\rho/\rho_0$,
$v'=v/v_0$, $B'=B/B_0$, $p'=p/p_0$.
       We also introduce the
dimensionless  time, $T=t/P_0$.
     Subsequently, the primes are dropped,
but the physical values can easily be restored for a particular
case.

Here, we discuss the numerical parameters for a typical T Tauri
star. We take the mass and radius of the star to be
    $M=0.8 M_\odot$ and $R_*=1.8 R_\odot$.
Our reference length is approximately equal to the inner radius of
the disk $R_0\approx 2.86 R_* \approx 3.6\times 10^{11}~{\rm cm}$.
         The reference  velocity  is
$v_0 \approx 1.93\times 10^7~{\rm cm/s}$, and the corresponding
time-scale is $t_0\approx R_0/v_0 \approx 1.89\times 10^4~{\rm
s}$.
      The size of the simulation region
corresponds to $R_{\rm max}=0.34~{\rm AU}$.
       The period of  Keplerian
rotation of the inner radius of the disk is $P_0 \approx 1.38~
{\rm days}$.

     Consider $B_*=10^3~{\rm G}$ at
the surface of the star. Then at $R=R_0$, the reference magnetic
field is  $B_0=B_*(R_*/R_0)^3 \approx 42.7 ~{\rm G}$ and the
reference magnetic moment is $\mu_0=B_0 R_0^3 \approx 2.0\times
10^{36}~{\rm G cm^{-3}}$.
          The reference density is
$\rho_0=4.89\times 10^{-12}~{\rm g/cm^3}$ or $n = 3.06\times
10^{12}~{\rm cm^{-3}}$ which is typical for T Tauri star disks.
The reference mass accretion rate is $\dot M_0 = \rho_0 v_0 R_0^2
\approx 1.2\times 10^{19}~{{\rm g/s}} \approx 1.92\times 10^{-7}~
{\rm M_\odot}/{\rm year}$.

\subsection{Initial and Boundary  Conditions}

       To derive the initial distribution
of density and pressure in the disk and corona, we use  {\it
quiescent} initial conditions which were developed for our
axisymmetric two-dimensional (R02) and test three-dimensional
simulations (K02).
    The main idea for creation of
quiescent initial conditions was: (1) to minimize the initial
magnetic braking between the slowly rotating dipole and matter of
the disk rotating with Keplerian velocity, and (2) to minimize
the effect of the rapid twisting of the magnetic field between the
disk and corona as a result of the fast turn on of the disk
rotation.
      The first process typically leads
to fast accretion of matter in the disk with a velocity close to
free-fall, while the second process leads to the formation of a
strong $B_\phi$ component near the disk and a transient matter
outburst to the corona (e.g., Hayashi et al. 1996; Miller \& Stone
1997).
      To minimize these factors, we
had the corona rotate with the Keplerian velocity of the disk
at  some distance $r > R_d$ from the axis.
      Thus, the
corona is differentially rotating with constant angular velocity
on cylinders of equal radius $r$, $\omega(r)=\rm const$ (see also
Romanova et al. 1998).

We establish the pressure balance on the boundary between the disk
and corona such that the disk below this boundary has high density
$\rho_d$ and low temperature $T_d$, while corona above this
boundary has low density $\rho_c << \rho_d$ and high temperature
$T_c >> T_d$. Then we derive the density and pressure distribution
in the whole region taking into account the differential rotation
of the disk and corona, and force balance between gravitational,
pressure, and centrifugal forces (see R02 for details). In the
typical density distribution, density in the disk gradually
increases outward, while density in the corona decreases while
moving to the axis (see Figure 1 of R02).
   We take the ratios:
$\rho_c/\rho_d=0.03 $ and $T_d/T_c=0.03$.
      Note, that in two-dimensional
simulations we took the ratios,  $\rho_c/\rho_d=0.01 $ and
$T_d/T_c=0.01$.
       We have chosen softer ratios for three-dimensional
simulations, in order to get a smaller Alfv\'en velocity
$v_A=B/\sqrt{4\pi\rho}$, and larger time-step $\Delta t\sim \Delta
R/v_A$. Test two-dimensional and three-dimensional simulations
with both sets of ratios have shown that in both cases the corona
does not influence the disk flow appreciably.

We place the initial inner radius of the disk $R_d$ at the
magnetospheric radius $r_m$, where the magnetic pressure of the
dipole field balances the matter pressure of the disk, $p+\rho
v^2={\bf B}^2/{8\pi}$. We chose $B_0$ at this point and $\rho_d$
such that $R_d=r_m=1$.
      This choice is based on the
fact that we often observe that at small inclination angles the
streams  always start from the magnetospheric radius $r_m$ (see
also Pringle \& Rees 1972; Ghosh \& Lamb 1979a,b for theoretical
base).
       This equilibrium
was checked numerically in two- and three-dimensional simulations
(see R02,  and \S 3.1 of this paper).

We suppose that the star rotates relatively slowly in the sense
that its corotation radius is at $r_{cor}=3$.
       This corresponds to
$\Omega_*=\sqrt{GM/r_{cor}^3}\approx 0.19=0.04 \Omega_{K*}$, where
$GM=1$ in our dimensionless units.
     For the case of T Tauri stars,
this corresponds to a  slow rotation rate ($T\approx 9.4~ {\rm
days}$ for the parameters used in \S 2.1 and in R02).
      Higher rotation rates including the ``propeller" regime
will be described in Romanova et al. (2004b).

\smallskip

      The {\it boundary conditions} are
similar to those used in K02.
           At the inner boundary
$R=R_*$, we take  ``free" boundary conditions ${\partial/\partial
R}=0$ for all variables.
       This  results in the
``absorption'' of the incoming matter so that there is no standoff
shock.
          Further, the inner boundary is treated as a
perfect conductor rotating at the rate ${\bf \Omega}=\Omega
\hat{\bf z}$.
     In the reference frame rotating with
the star the flow velocity  is parallel to ${\bf B}$ at $R=R_*$.
           The boundary
condition at $R_*$ on the magnetic field has ${\partial(R
B_\phi)}/{\partial R}=0$.
       At the outer boundary $R=R_{\rm max}$,
free boundary conditions are taken for all variables.
       We investigated the possible influence of the outer
boundary conditions by running cases where the simulation regions
had $R_{\rm max}=4.9$, $14$, and $68$.
      We find the same results except for the case
$R_{\rm max}=4.9$ where the accretion rate decreases too fast
because the reservoir of matter in the disk is too small.
      Results for
the medium and large regions are very close, so that we take
$R_{\rm max}=14$ as a standard size of the simulation region.

\subsection{The Grid and Parallelization of the Code}

       As mentioned the inner boundary of the simulation
region (the ``star") is taken to be at $R_*=0.35$. The size of
external boundary depends on the grid (see below).

         The spherical grid was inhomogeneous
in the $R-$direction.
        The inhomogeneity
was such that cells at any distance $R$ were  approximately square
with $\Delta R \sim R \Delta \theta$ for fixed $\Delta \theta$.
         This grid gives a high space
resolution close to the star which is important in this problem.
At the same time, simulations in a very large region may be
performed. The inhomogeneous grid is a smooth analog of nested
grids used for example by  GBW99.

The angular resolution was taken to be $N_\theta=29$ for each
block in most cases, and $N_\theta=41$ in test cases.
      The number
of points in $R-$ direction determines the size of the simulation
region.
      The main simulations were done with the radial grid  $N_R=70$
which corresponds to $R_{\rm max}=14 \approx 40 R_*\approx 0.34
~{\rm AU}$. Simulations at the larger grid $N_R=96$ corresponding
to $R_{\rm max}=68\approx 194 R_*\approx 1.6~{\rm AU}$ were also
performed for testing the external boundary conditions.

Note, that each of the six blocks consists of $70\times 29\times
29$ cells.
      In comparison with a spherical coordinate system,
this grid corresponds to  $N_R\times N_\theta\times
N_\phi=70\times 58\times 116$.
     This grid gives high resolution in the vicinity of the star
and inner region of the disk.
      For the typical grid $N_\theta=29$,
the size of the smallest cells
    near the surface of the star is
$\Delta R = 0.019\approx 0.053 R_*$.
      This cell size is $1.7$ times larger
than in our two-dimensional simulations of the corresponding
problem (R02), but about $1.8$ times smaller than  cells in the
innermost grid of GBW99 in their two-dimensional simulations, who
had resolution $\Delta R \approx 0.1 R_*$ in the innermost grid.
        At larger distances from
the star, the grid size becomes larger but at the magnetospheric
distance $R=1$ it is still very small, $\Delta R\approx 0.1 R_*$,
and continues to be sufficiently small in the inner region of the
disk $R \lesssim 5$ which is the most important for our current
investigation.
    At the external boundary, $R_{max}=14$,
    $\Delta R = 2 R_*$, which is
small compared with the disk thickness at this distance.
      These numbers show that this
resolution and grid are satisfactory for three-dimensional
simulations of the matter flow around the magnetized star.
      Note, that in K02 it was shown that
even sparser grids such as $26\times 15\times 15$ give
satisfactory results.

Fast runs of the three-dimensional code at high grid resolutions
become possible due to {\it parallelization} of the code.
      First, the code was
naturally parallelized to 6 processors, where each of six blocks
was assigned to one of six processors.
        Next, we divided each block into equidistant layers
$N_{layers}$ in $R-$ direction. Division onto
$N_{layers}=2,3,4,...$ etc.  allowed us to use $N_{proc}=6\times
N_{layers}=12, 18, 24, ...$ number of processors.
      We typically used $N_{layers}=8$, that is,
$48$ processors. Simulations  with larger numbers of processors
were also possible and still produced an increase in the
computation speed, but the efficiency of using the parallel
machine at $N_{proc} > 48$ declined.

       The reported simulations were performed mainly
on the fastest V2 nodes of the ``Velocity Cluster'' at the Cornell
Theory Center. Each of the V2 nodes is a dual processor  $2.4$ GHz
Pentium 4  computer.
       Typical simulations of $T=10 P_0$ rotations with the
grid $70\times 29\times 29$ for each of $6$  blocks  took
approximately 40-50 wall-clock hours on the 48 processors of V2
nodes.
      This was a reasonable time of simulations in that
it allowed us to investigate the three-dimensional flows at
different inclination angles (this paper) and other parameters
(Romanova et al. 2004b), and  to perform  test runs with the grid
$70\times 41\times 41$.

\section{Test Numerical Simulations}

First, we discuss tests of the code with the above mentioned
initial and boundary conditions in  hydrodynamic simulations, that
is, $B=0$.
     Secondly, we discuss tests of the code for
accretion to a rotating star with an aligned dipole magnetic
field, $\Theta=0$. Thirdly, we  compare  results from our
three-dimensional simulations with analogous two-dimensional
axisymmetric simulations.

%%%%%%%%%%%%%%%%%%%%%%%%%%%%%%%%%%%%%%%%%%%%%%%%%%%%%%%%%%%%
\begin{figure*}[t]
\epsscale{1.6} \plotone{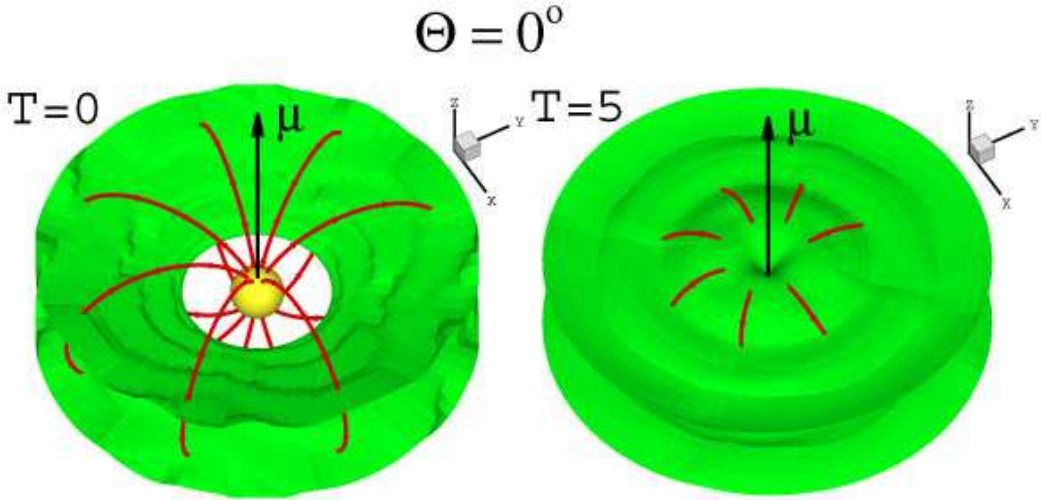} \caption{The figure shows
results of three-dimensional MHD simulations of disk accretion to
an aligned rotator  ($\Theta=0$) for $T=0$ and $T=5$. Time $T$ is
measured in units of Keplerian rotation period at the inner radius
of the disk $r=R_d=1$.
  The isodensity
contours (background) are shown at $\rho=0.15$. The solid lines
represent the magnetic field lines. The grid $70\times 29\times
29$ is taken for each of six sectors of the cubed sphere grid.
Only part of the simulation region $R \leq 3$ is shown.}
\label{Figure 2}
\end{figure*}
%%%%%%%%%%%%%%%%%%%%%%%%%%%%%%%%%%%%%%%%%%%%%%%%%%%%%%%%%%%%

\subsection{Hydrodynamic Simulations ($B=0$)}

       Hydrodynamic simulations are important
for checking the initial setup of the problem.
       Simulations were
done for grids $N_1=70\times 29\times 29$ and $N_2=70\times
41\times 41$ for each of six blocks.
      The first grid was the
base grid for  subsequent runs with the magnetic field, while the
second grid was used for comparisons.
      Simulations with both grids
have shown that the disk-star system stays in  equilibrium during
many rotations $T > 15 P_0$, and only slightly evolves due to
numerical viscosity. Inward velocity associated with numerical
viscosity is very small, $v\approx 0.001-0.003 v_{Kd}$, where
$v_{Kd}=\sqrt{GM/R_d}$ is the Keplerian velocity at the distance
$r=R_d$.
      For longer times, the outer regions of the
disk become thicker because the grid resolution in that region is
relatively low.
      No high  velocities were
observed in the corona.
      The initial
differential rotation in the corona was gradually disturbed, and
the corona become almost homogeneous but still low-density.
       The accretion
rate to the star was calculated to be $\dot M = 0.01 - 0.02$ in
the dimensionless units (see \S 2.2) for both grids $N_1$ and
$N_2$.
      Note that the accretion rate
obtained subsequently in the magnetic cases is about $10$ times
larger.

Thus, hydrodynamic simulations have shown that the disk-star
system stays in the gravity/pressure/centrifugal equilibrium for a
sufficiently long time. In three-dimensional simulations we were
able to follow the evolution during $5-12$ rotations $P_0$.
Compared to our two-dimensional simulations (R02) which were
followed until $T > 50$ rotation, here, we do not have viscosity
incorporated into the code. Longer simulations with incorporated
viscosity will be done in the future research.

In 2D simulations we noticed, that the magnetospheric funnel flow
establishes during one rotation period of the inner disk.
Subsequently, the flow changes insignificantly during the
subsequent $50$ rotations and variation depends on the incoming
matter flux. In simulations of the accretion to the inclined
rotator, we observed that the flow  also  establishes during the
first several rotations, so that as presented in this paper
simulations which lasted $5-12$ rotations are valuable for
understanding of 3D funnel flows.

\subsection{Accretion to an Aligned Rotator, $\Theta=0$}

Next, we performed three-dimensional simulations of accretion to a
magnetized star.
     First, we chose the case where the
magnetic axis is perpendicular to the disk and  $\Theta=0$.
      We took the standard parameters for
initial and boundary conditions as described in \S 2.1.
      We observed that during the first two
rotation periods, matter of the disk moved inward, but later it
stopped at the magnetospheric radius $r_m\approx 1 - 1.2$ and went
up/down out the disk plane, forming magnetospheric funnel flows.
The flow is cylindrically-symmetric, as it should be. The shape of
the funnel flow is similar starting from  $T \approx 3$ and during
larger $T$.
       Figure 2 shows  magnetospheric flow at $T=0$ and $T=5$.
Here and in subsequent plots we show only the inner part of the
simulation region, $R \le 3$, in order to resolve the funnel flow
and the inner region of the disk.
       The magnetospheric flow at $\Theta=0$ has a
relatively low density, so that we show isocontours  at the
density level $\rho=0.15$.

The star rotates slowly, $\Omega_*=0.04 \Omega_{*K}$ with the
co-rotation radius  at $r_{cor}=3$. Thus, the  magnetic field
lines threading the disk at $r \lesssim 3$ have an angular
velocity lower than the Keplerian, and they transfer negative
angular momentum to the disk matter. This leads  to the slow
inward flow of matter to the star with velocity $v\approx
0.03-0.05 v_{Kd}$.
      This  flow results from the magnetic braking,
      but the velocity of the flow
is much smaller than that obtained using non-equilibrium initial
conditions (e.g., Hayashi et al. 1996).

      To understand the behavior of
the magnetic field, we chose a set of magnetic field lines which
initially threaded the disk  at a distance $r\approx 2.8$, and
show the same set of lines at subsequent moments of time.
      One can see that accreting
matter   drags these field lines inward. This demonstrates that
magnetic field lines are well frozen to the disk. They rotate with
the disk in the azimuthal direction and slowly move inward in the
radial direction.
    No opening of magnetic field lines was
observed, because the corona right above the disk is
matter-dominated (e.g., Romanova et al. 1998).

We should note that the boundary between the disk and
magnetosphere of the aligned dipole is a possible site for
  onset of 3D instabilities, such as the Relegh-Taylor instability
  which may lead to direct accretion of matter through
  magnetosphere to the star
  (see Arons \& Lea 1976a,b, Scharlemann 1978). No accretion
  through 3D instabilities was observed in our simulations. Possibly special
  simulations are needed with even higher resolution.

%%%%%%%%%%%%%%%%%%%%%%%%%%%%%%%%%%%%%%%%%%%%%%%%%%%%%%%%%%%%
\begin{figure*}[t]
\epsscale{1.0} \plotone{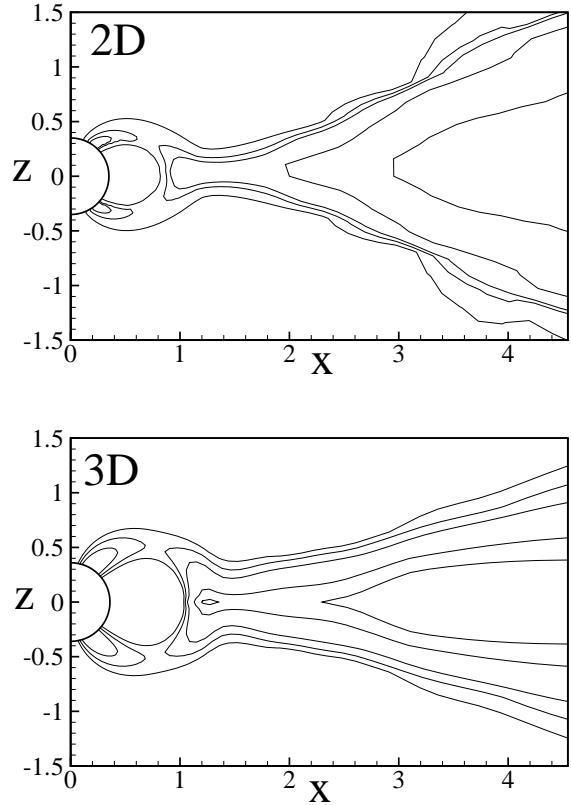} \caption{Comparison of the
density profiles for disk accretion to an aligned rotator obtained
from two-dimensional, axisymmetric simulations in spherical
coordinates  (R02) (top panel), and from three-dimensional
simulations in the cubed-sphere coordinates (bottom panel), both
at $T=5$.} \label{Figure 3}
\end{figure*}
%%%%%%%%%%%%%%%%%%%%%%%%%%%%%%%%%%%%%%%%%%%%%%%%%%%%%%%%%%%%
%%%%%%%%%%%%%%%%%%%%%%%%%%%%%%%%%%%%%%%%%%%%%%%%%%%%%%%
\begin{figure*}[t]
\epsscale{1.6} \plotone{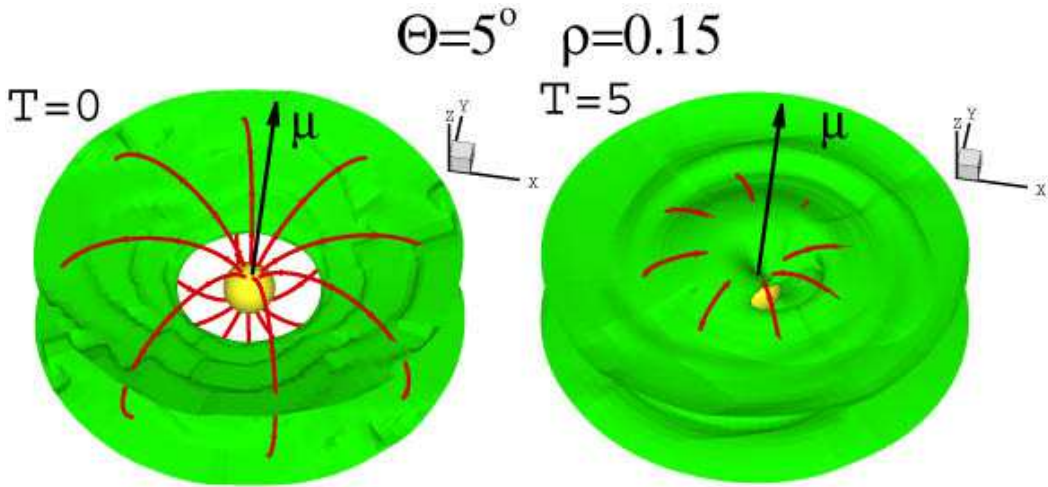} \caption{Simulations of
accretion to an inclined dipole rotator for an inclination angle
$\Theta=5^\circ$.
      Background represents density
at the level $\rho=0.15$.
       Time $T$ is measured in periods of
Keplerian rotation at $R=1$.} \label{Figure 4}
\end{figure*}
%%%%%%%%%%%%%%%%%%%%%%%%%%%%%%%%%%%%%%%%%%%%%%%%%%%%%%%

%%%%%%%%%%%%%%%%%%%%%%%%%%%%%%%%%%%%%%%%%%%%%%%%%%%%%%%
\begin{figure*}[t]
\epsscale{1.6} \plotone{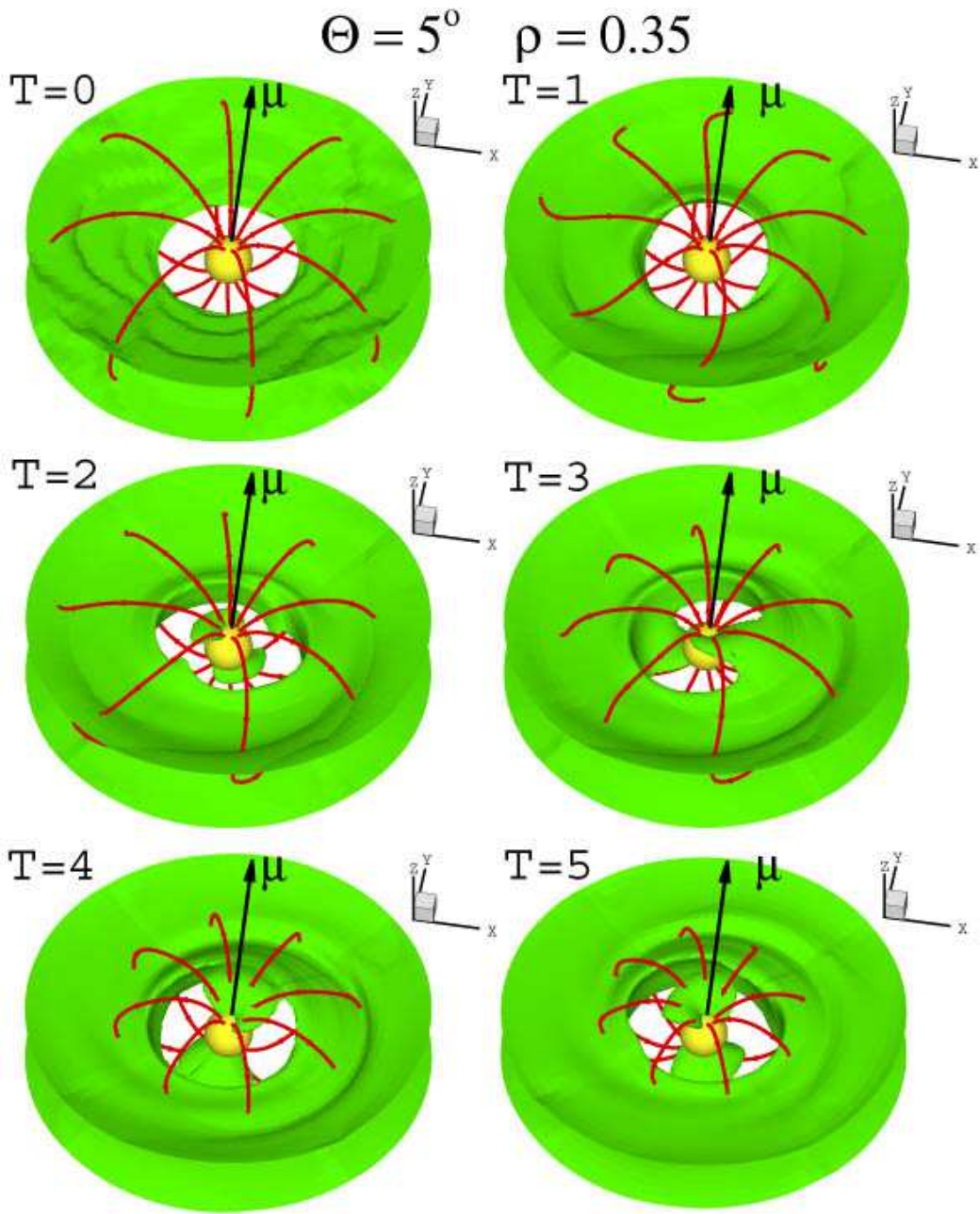} \caption{Simulations of disk
accretion to a misaligned dipole rotator for the case where the
dipole moment is inclined by  an angle $\Theta=5^\circ$. The
background represents the density surface where $\rho=0.35$. Time
$T$ is measured in periods of Keplerian rotation at $R=1$.}
\label{Figure 5}
\end{figure*}
%%%%%%%%%%%%%%%%%%%%%%%%%%%%%%%%%%%%%%%%%%%%%%%%%%%%%%%

%%%%%%%%%%%%%%%%%%%%%%%%%%%%%%%%%%%%%%%%%%%%%%%%%%%%%%%
\begin{figure*}[t]
\epsscale{1.6} \plotone{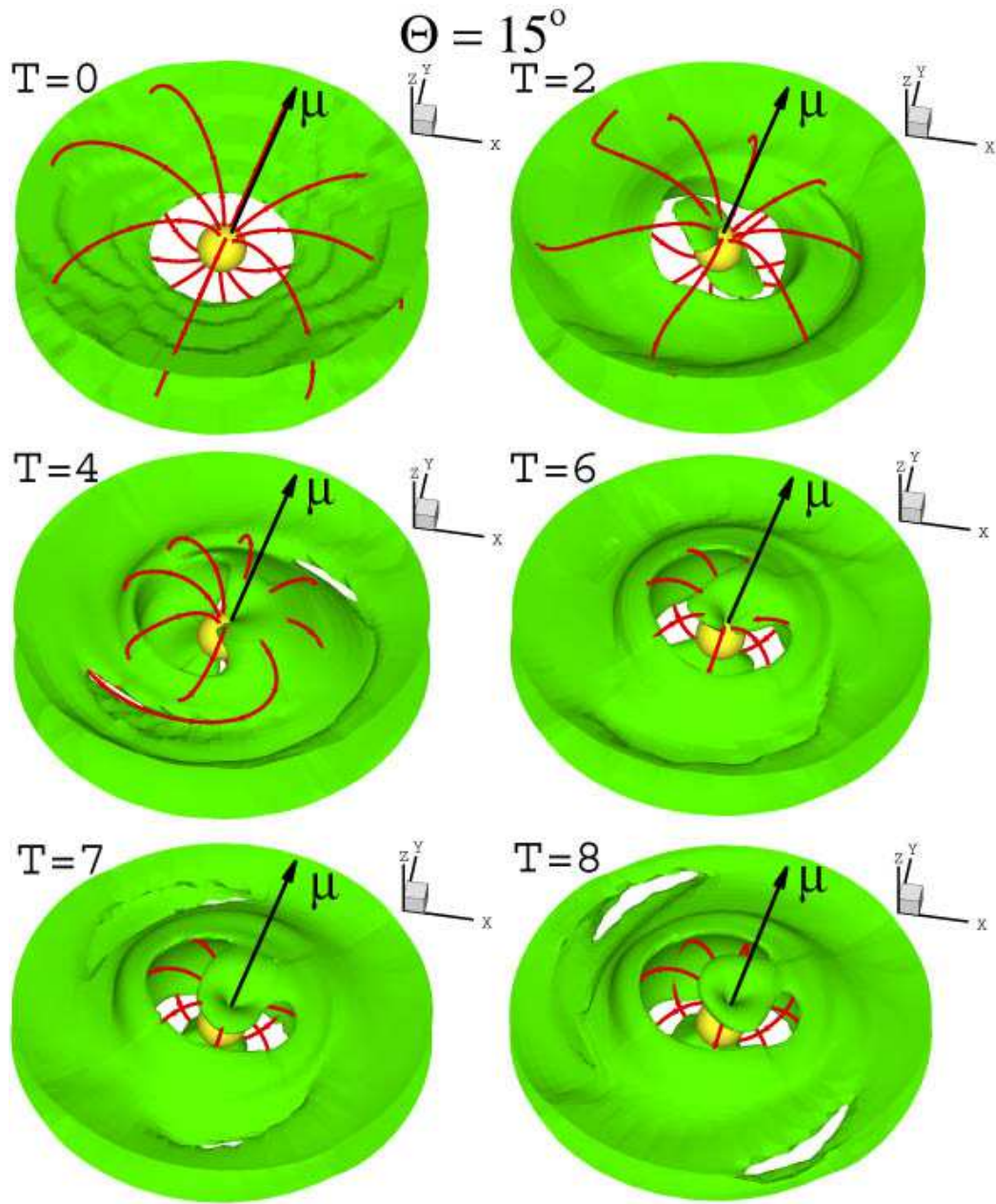} \caption{Matter flow around an
inclinded dipole rotator with the dipole  inclination angle
$\Theta=15^\circ$.
     The background
represents the density isocontours at $\rho=0.35$.}
\label{Figure6}
\end{figure*}

%%%%%%%%%%%%%%%%%%%%%%%%%%%%%%%%%%%%%%%%%%%%%%%%%%%%%%%
\begin{figure*}[t]
\epsscale{1.6} \plotone{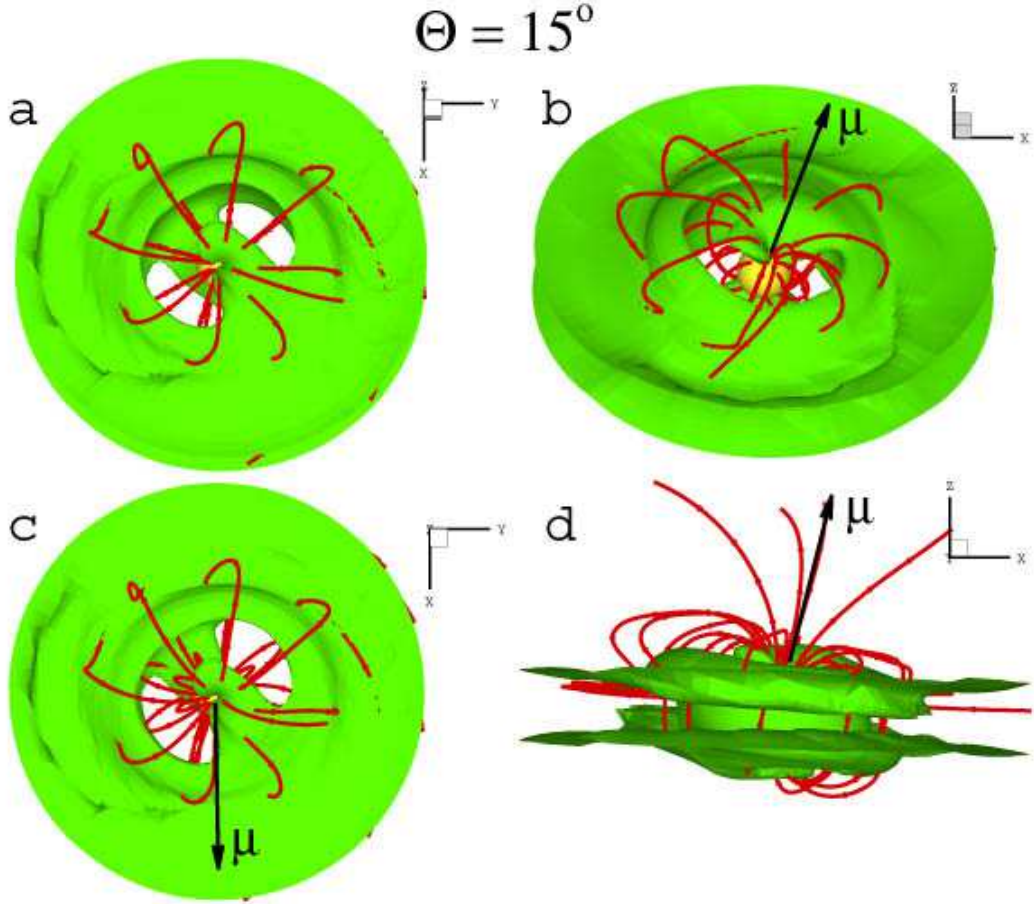} \caption{The figure shows
different projections  of the flow for an inclination angle
$\Theta=15^\circ$ at time $T=6$.
     The background represents the density at  $\rho=0.35$; the
solid lines are magnetic field lines. In panel  (a) the $\mu$
vector is directed towards the observer; in panel (b) the
orientation is the same as in Figure 6; panel (c) shows  the view
    along the $Z$ axis with
    vector $\rvecmu$ directed downward; panel
    (d) shows the side view
    along the $Y$ axis.}
\label{Figure 7}
\end{figure*}
%%%%%%%%%%%%%%%%%%%%%%%%%%%%%%%%%%%%%%%%%%%%%%%%%%%%%%%

%%%%%%%%%%%%%%%%%%%%%%%%%%%%%%%%%%%%%%%%%%%%%%%%%%%%%%%
\begin{figure*}[t]
\epsscale{1.6} \plotone{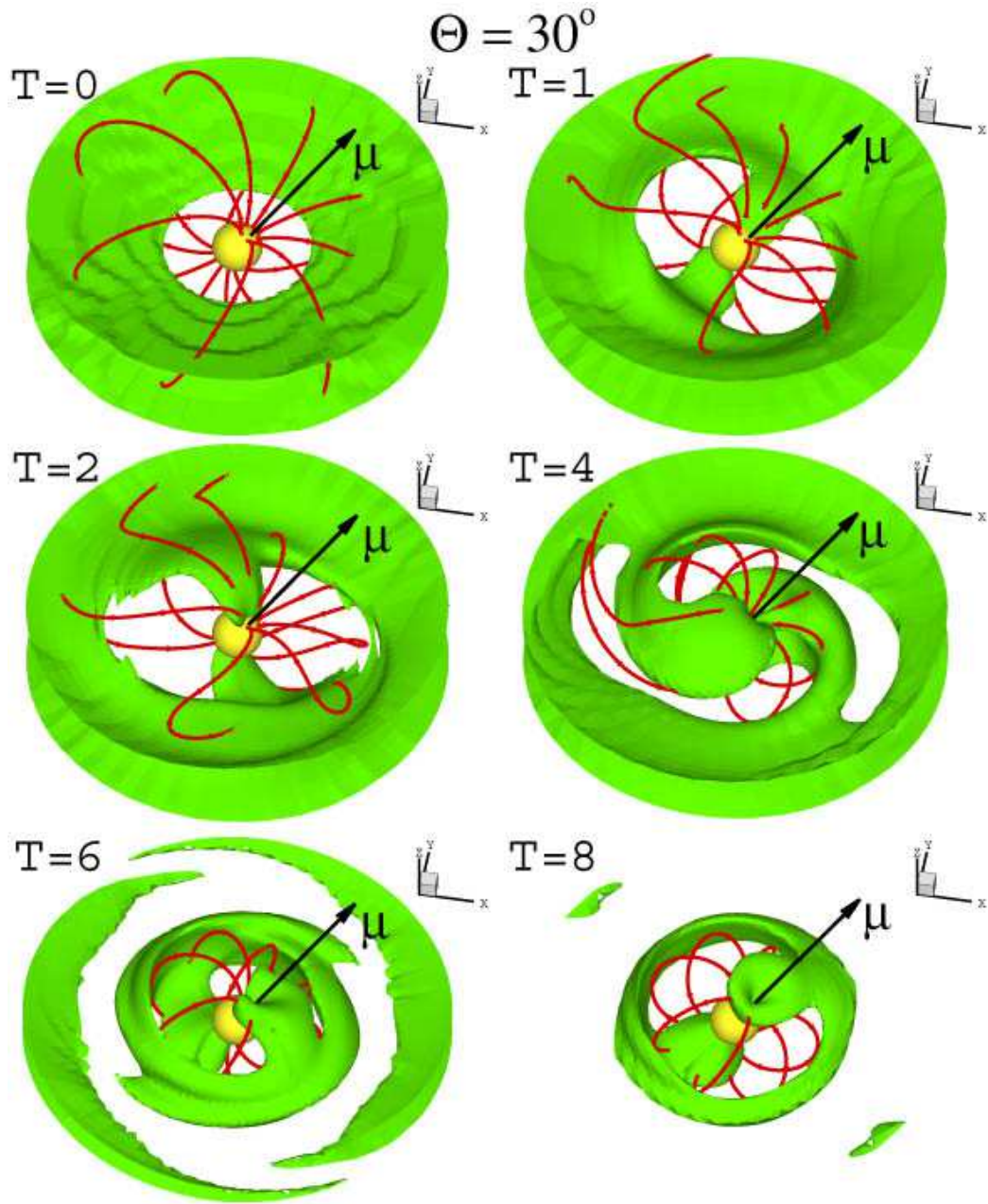} \caption{Matter flow around the
dipole rotator for an inclination angle $\Theta=30^\circ$. The
background represents the density isocontours at $\rho=0.4$.}
\label{Figure 8}
\end{figure*}
%%%%%%%%%%%%%%%%%%%%%%%%%%%%%%%%%%%%%%%%%%%%%%%%%%%%%%%

%%%%%%%%%%%%%%%%%%%%%%%%%%%%%%%%%%%%%%%%%%%%%%%%%%%%%%%
\begin{figure*}[t]
\epsscale{1.6} \plotone{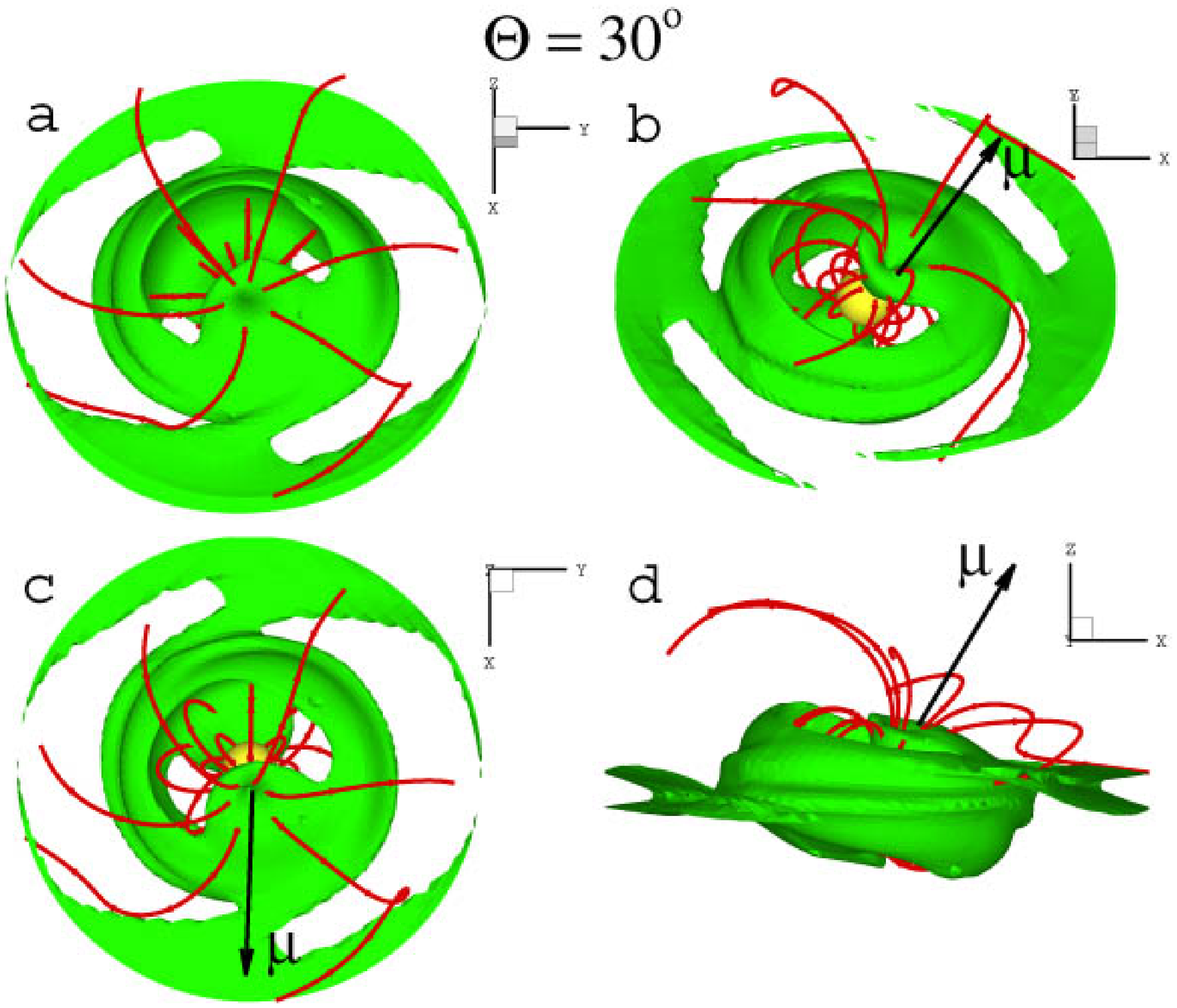} \caption{The figure shows
different projections  for an inclination angle $\Theta=30^\circ$
and time $T=7$.
      The background represents density
at the level $\rho=0.35$, and the solid lines are magnetic field
lines.
     In panel  (a) the magnetic moment $\rvecmu$
is directed towards the observer; in panel (b) the orientation is
the same as in Figure 8; panel (c) shows the view
    along the $Z$ axis with
    $\rvecmu$ directed downward;
and panel
    (d) shows the side view
    along the $Y$ axis.}
    \label{Figure 9}
\end{figure*}
%%%%%%%%%%%%%%%%%%%%%%%%%%%%%%%%%%%%%%%%%%%%%%%%%%%%%%%

\subsection{Comparison of Two-dimensional and Three-dimensional
Simulations}

We compared results of axisymmetric three-dimensional simulations
with corresponding  two-dimensional simulations.
      Two-dimensional simulations were done in
spherical coordinates with the grid
    $N_R\times N_\theta =70 \times 58$
which is the same as  the cubed sphere grid $N_R\times N\times N =
70\times 29\times 29$ for each block.
%      We found that the
%results are  pretty close to each other in spite of the difference
%in the space dimension and  the type of the coordinate system
%used.
        Figure 3 shows the meridional
cross-sections of matter flow for both cases after $T=5$
rotations.
       Only the inner region $R < 4.5$ is
shown in order to resolve the matter flow around the
magnetosphere.
      One can
see that results of two-dimensional and three-dimensional
simulations are qualitatively similar. However, in 3D simulations
the magnetosphere radius is $\sim 10-20\%$ larger than in 2D case:
 the funnel flow starts at $r_m\approx
0.9-1.1$ in the two-dimensional simulations and at $r_m\approx
1.0-1.2$ in the three-dimensional simulations. Besides, the funnel
flow in 3D simulations forms later, after two rotations, compared
to one rotation in 2D simulations (R02). These differences seems
to be connected with the difference in the accretion rate.
Accretion rate  in 3D case is about $\sim 10-20\%$ smaller than in
2D case: $\dot M_{3D} \approx 0.11-0.13$ versus $\dot M_{2D}
\approx 0.12-0.16$.  In  present simulations the inward accretion
of matter occurs as a result of small initial magnetic braking
%which is associated with small initial non-stationarity of the
%system
(see  \S 2.2).
 The magnetic braking
may have a different nature in 2D and 3D cases which may possibly
lead to the observed difference between results in 2D and 3D
simulations. In 2D simulations this braking is possibly stronger,
which leads to the enhanced accretion compared to 3D simulations.

Here, we should point out that magnetospheric structure does
depend on  the accretion rate, but {\it does not} depend on the
mechanism of accretion and angular momentum transport. In R02 and
in recent 3D simulations (R03), we introduced the simplified
$\alpha-$ type viscosity term to the equations, which permitted us
to vary the accretion rate. In real accretion disks the transport
of the angular momentum is connected most probably with
magneto-rotational instability (e.g., Balbus \& Hawley 1991,
1998).  Numerical investigation of this instability needs high
space resolution in the disk and typically only part of the disk
is considered (e.g., Hawley, Gammie \& Balbus 1995; Brandenburg et
al. 1995; Stone, et al. 1996; Fleming, Stone, \& Hawley 2000). In
the present simulations the high grid resolution was set in the
vicinity of the dipole, while resolution father out in the disk
was not sufficient to resolve this instability.

      These tests provide assurance about
the reliability of our three-dimensional MHD code. It is important
that at $\Theta=0^\circ$ the code supports axisymmetry of the
flow.  We next investigate disk accretion to an inclined dipole
rotator.

%%%%%%%%%%%%%%%%%%%%%%%%%%%%%%%%%%%%%%%%%%%%%%%%%%%%%%%%
\begin{figure*}[t]
\epsscale{1.6} \plotone{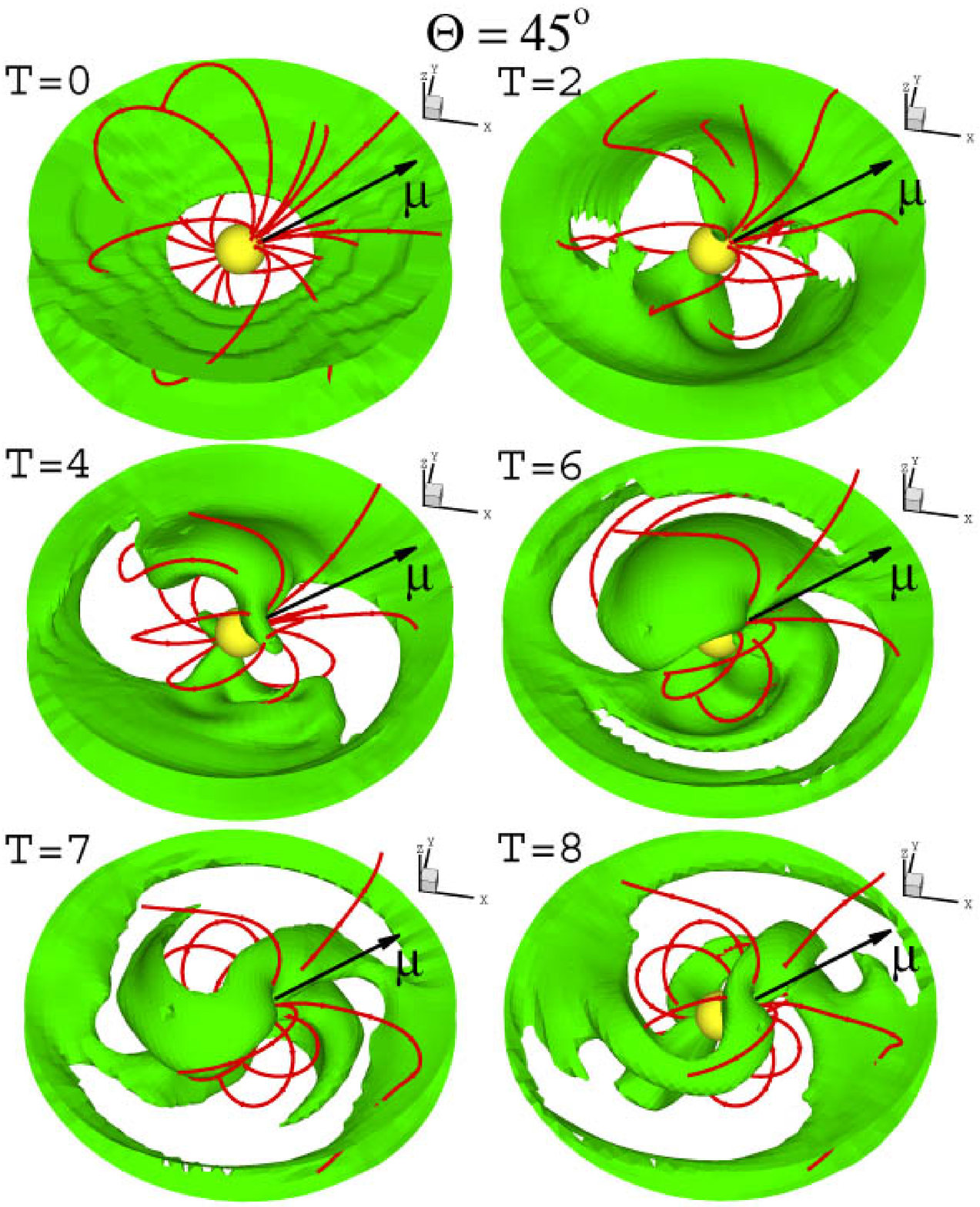} \caption{Matter flow to an
inclined rotator for an inclination angle $\Theta=45^\circ$ at the
times indicated.
      The background represents the
density surface $\rho=0.35$.} \label{Figure 10}
\end{figure*}
%%%%%%%%%%%%%%%%%%%%%%%%%%%%%%%%%%%%%%%%%%%%%%%%%%%%%%%%

%%%%%%%%%%%%%%%%%%%%%%%%%%%%%%%%%%%%%%%%%%%%%%%%%%%%%%%%
\begin{figure*}[t]
\epsscale{1.6} \plotone{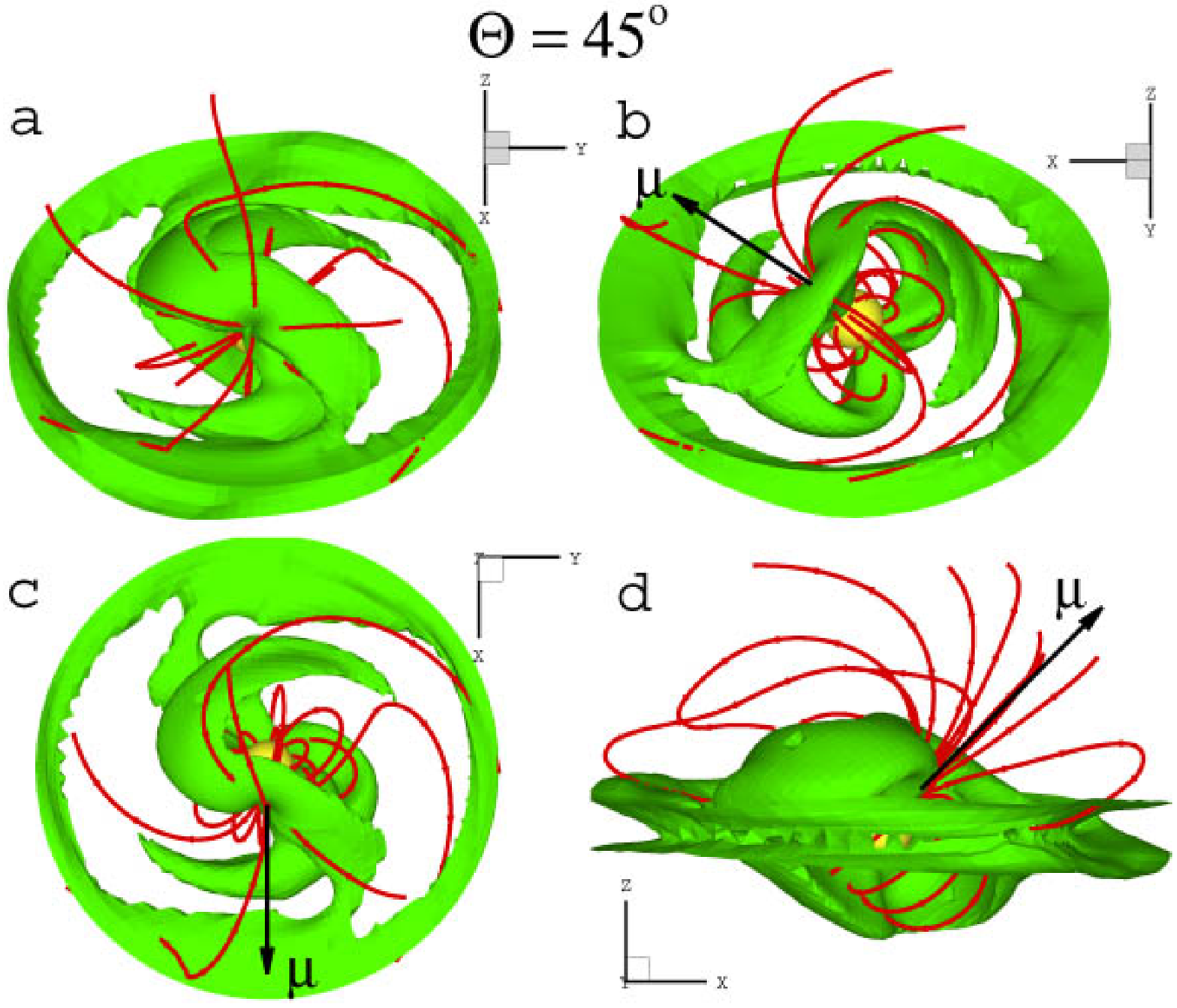} \caption{Different projections
of the accretion flow for  inclination angle $\Theta=45^\circ$ and
time $T=7$.
      The background represents the density surface $\rho=0.35$, and
the solid lines are magnetic field lines.
     In panel  (a) the magnetic moment $\rvecmu$
is directed towards the observer.
     Panel (b) shows one
    interesting projection; panel (c)
shows the  view along the $Z$ axis with
    $\rvecmu$ directed downward,  and panel
    (d) shows the the side view
    along the $Y-$axis.} \label{Figure 11}
\end{figure*}
%%%%%%%%%%%%%%%%%%%%%%%%%%%%%%%%%%%%%%%%%%%%%%%%%%%%%%%%

%%%%%%%%%%%%%%%%%%%%%%%%%%%%%%%%%%%%%%%%%%%%%%%%%%%%%%%%
\begin{figure*}[t]
\epsscale{1.6} \plotone{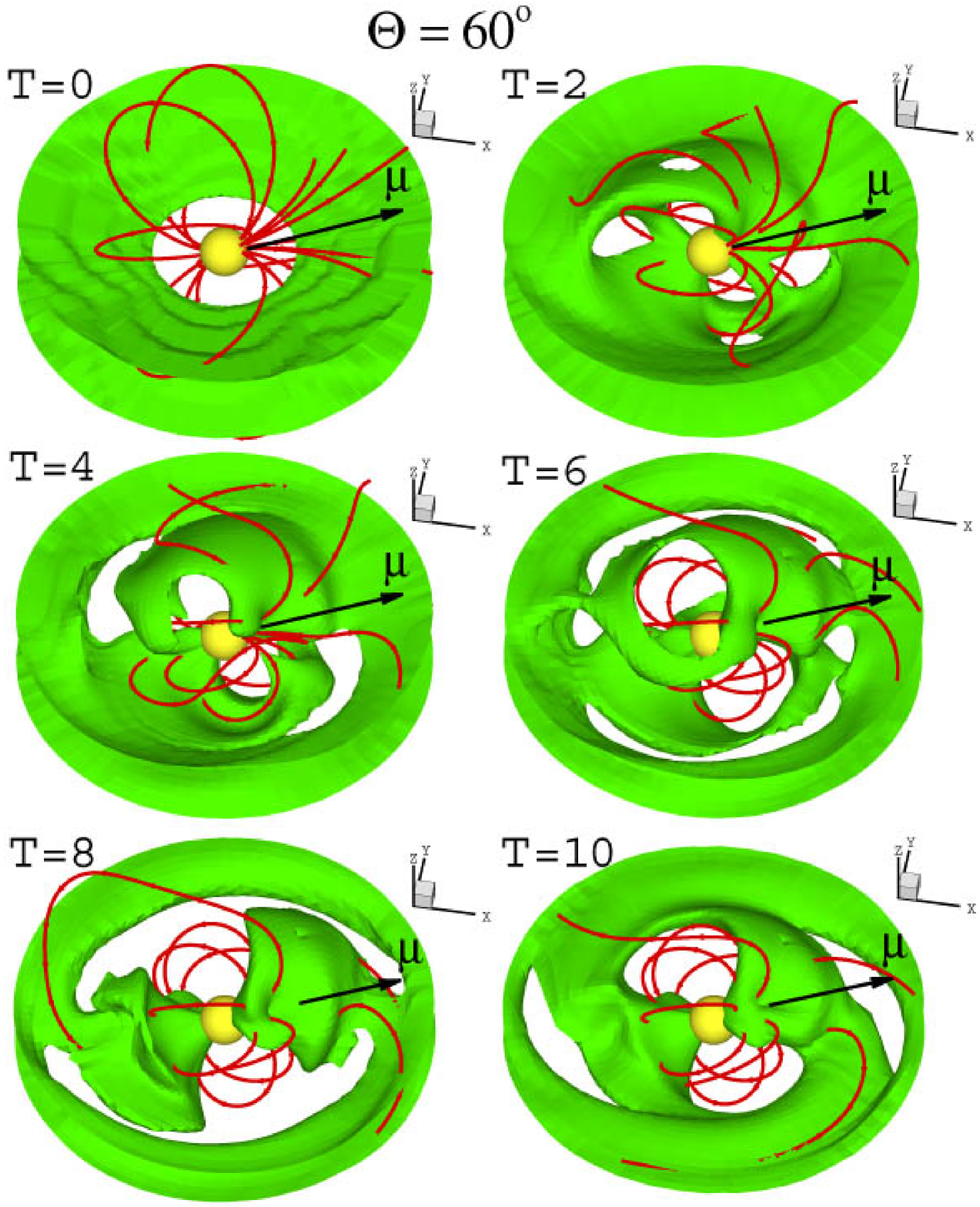} \caption{Accretion flow to an
inclined dipole rotator for an inclination angle
$\Theta=60^\circ$.
      The background
represents the density surface $\rho=0.35$.} \label{Figure12}
\end{figure*}
%%%%%%%%%%%%%%%%%%%%%%%%%%%%%%%%%%%%%%%%%%%%%%%%%%%%%%%%

%%%%%%%%%%%%%%%%%%%%%%%%%%%%%%%%%%%%%%%%%%%%%%%%%%%%%%%%
\begin{figure*}[t]
\epsscale{1.6} \plotone{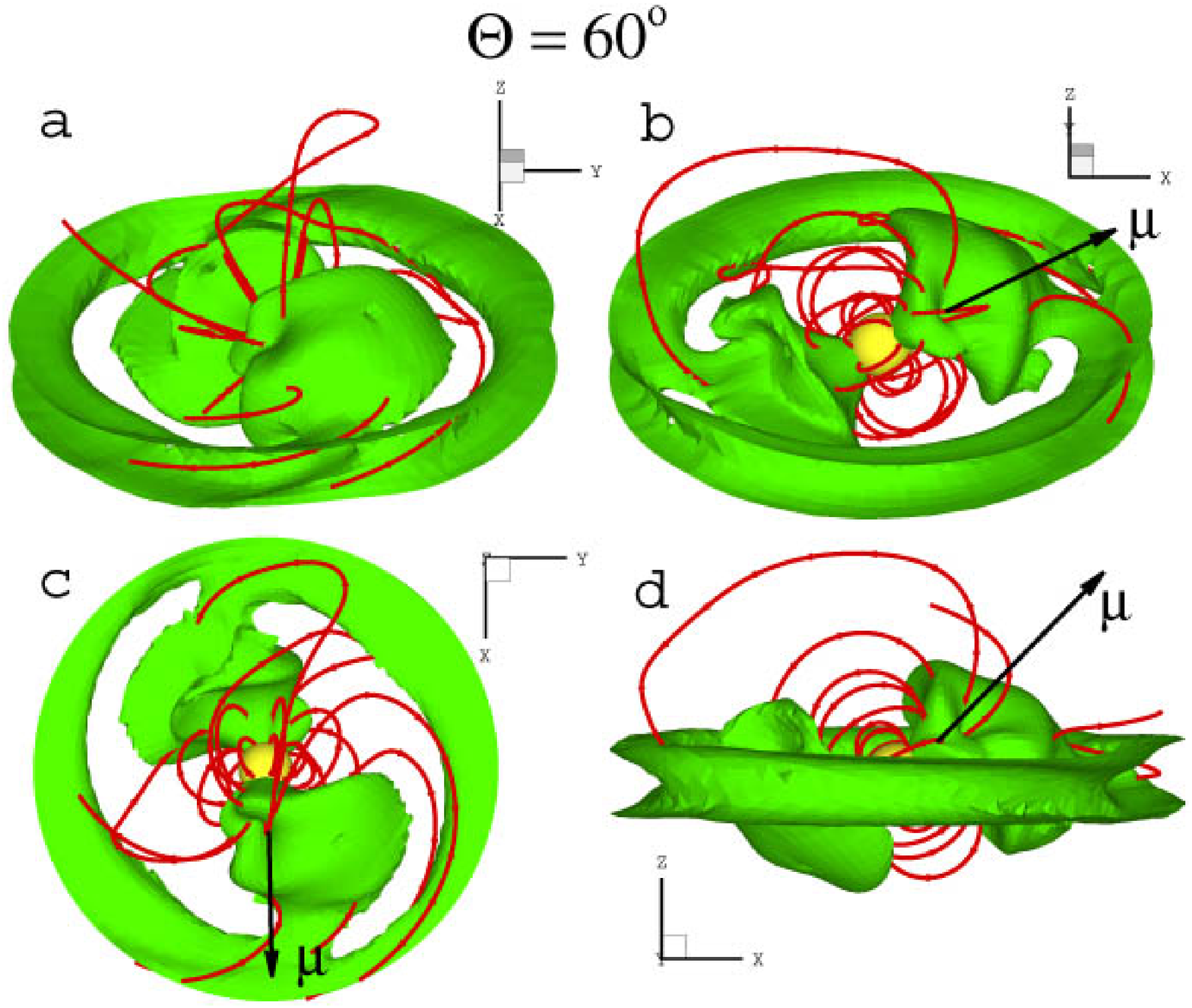} \caption{ Different projections
of the accretion flow for an inclination angle $\Theta=60^\circ$
at time $T=8$.
      The background represents the density surface
$\rho=0.35$; the solid lines are magnetic field lines. In panel
(a) the magnetic moment $\rvecmu$ is directed towards the
observer;  panel (b) shows one of the
    interesting projections; panel (c)
shows  view along the $Z$ axis with $\rvecmu$ directed downward;
and panel
    (d) shows the the side view
    along the $Y-$axis.}
    \label{Figure 13}
\end{figure*}
%%%%%%%%%%%%%%%%%%%%%%%%%%%%%%%%%%%%%%%%%%%%%%%%%%%%%%%%

%%%%%%%%%%%%%%%%%%%%%%%%%%%%%%%%%%%%%%%%%%%%%%%%%%%%%%%%
\begin{figure*}[t]
\epsscale{1.6} \plotone{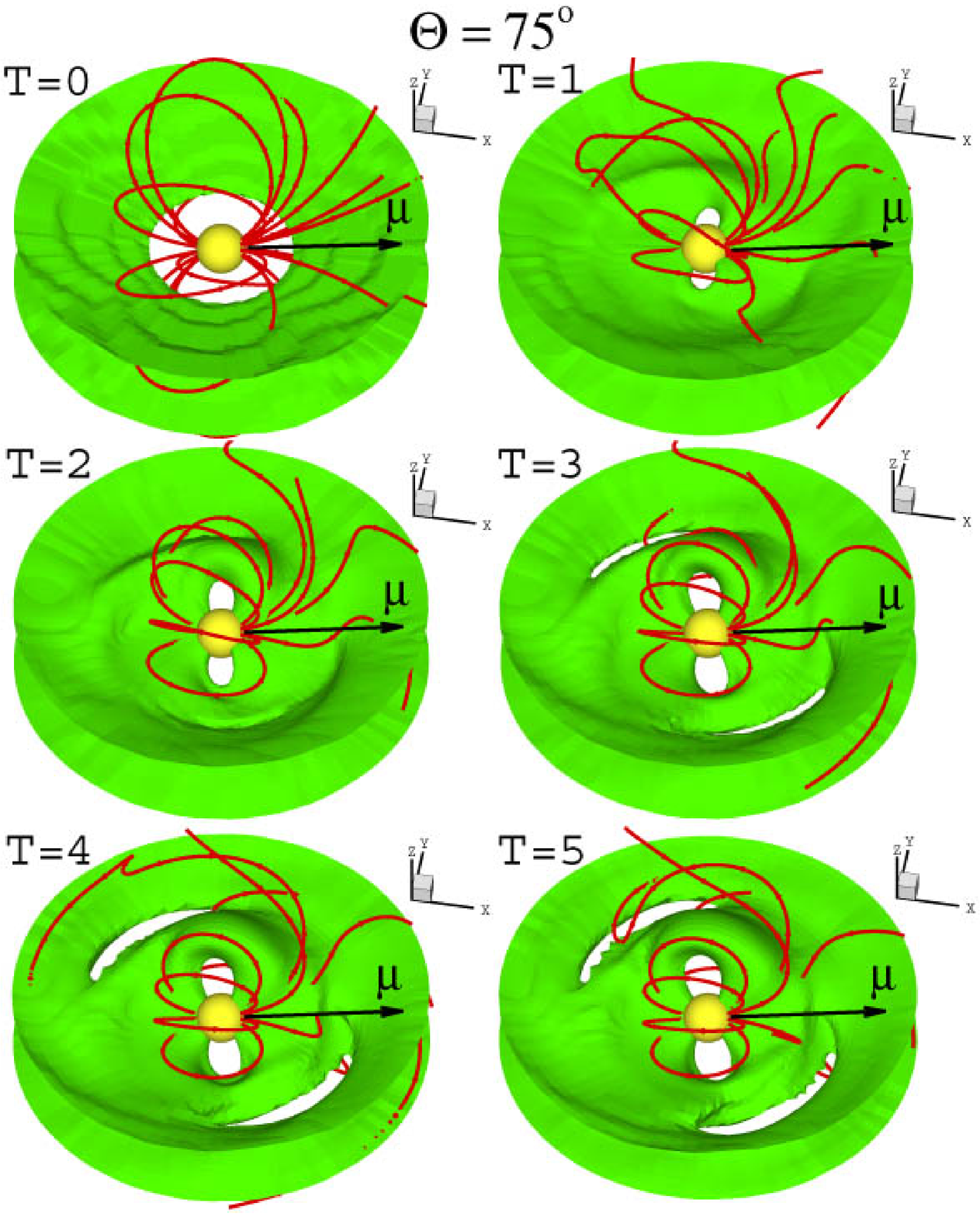} \caption{Accretion flow to an
inclined dipole rotator for an inclination angle
$\Theta=75^\circ$. The background represents the density surface
$\rho=0.45$.} \label{Figure14}
\end{figure*}
%%%%%%%%%%%%%%%%%%%%%%%%%%%%%%%%%%%%%%%%%%%%%%%%%%%%%%%%

%%%%%%%%%%%%%%%%%%%%%%%%%%%%%%%%%%%%%%%%%%%%%%%%%%%%%%%%
\begin{figure*}[t]
\epsscale{1.6} \plotone{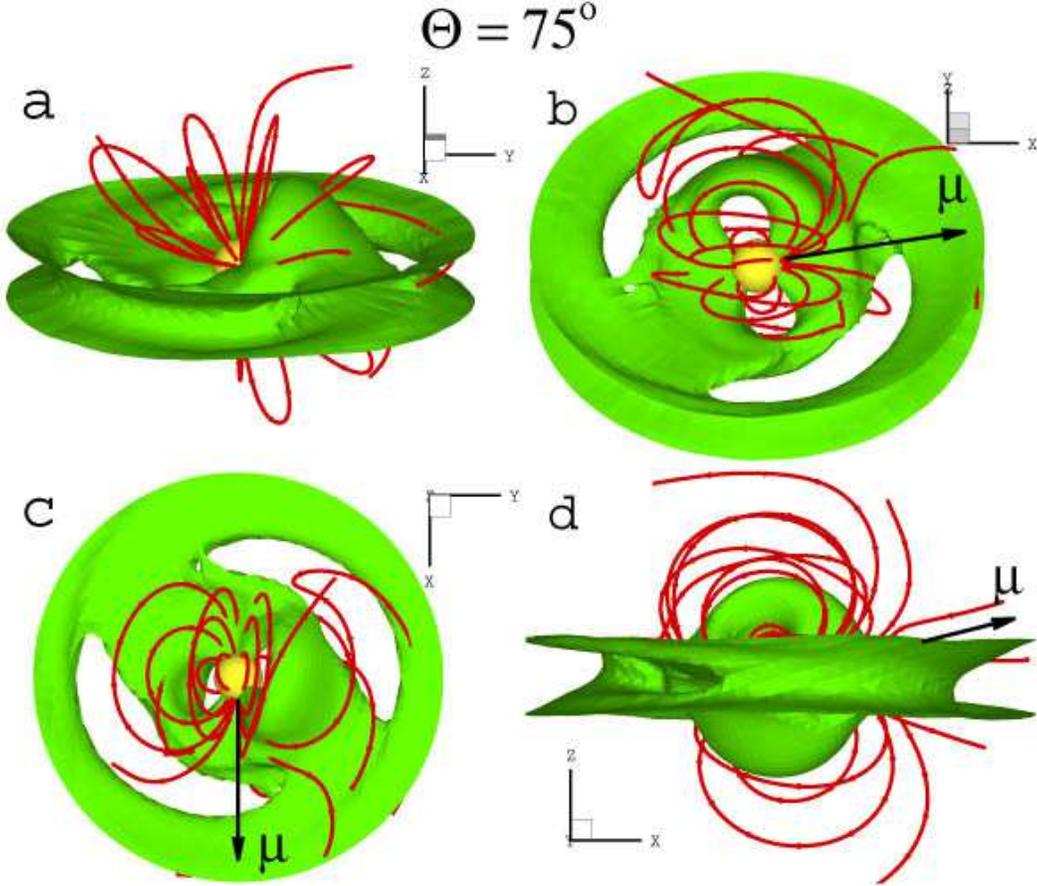} \caption{Different projections
of the accretion flow for an inclination angle $\Theta=75^\circ$
and time $T=5$.
      The background represents the density surface $\rho=0.45$;
the solid lines are magnetic field lines. In panel  (a) the
magnetic moment $\rvecmu$
    is directed towards the observer;  panel (b) shows a projection
    similar to one in Figure 16;
panel (c) shows the view along the $Z$ axis with $\rvecmu$
directed downward;and panel
    (d) shows the the side view
    along the $Y-$axis.}
\label{Figure 15}
\end{figure*}
%%%%%%%%%%%%%%%%%%%%%%%%%%%%%%%%%%%%%%%%%%%%%%%%%%%%%%%%

%%%%%%%%%%%%%%%%%%%%%%%%%%%%%%%%%%%%%%%%%%%%%%%%%%%%%%%%
\begin{figure*}[t]
\epsscale{1.5} \plotone{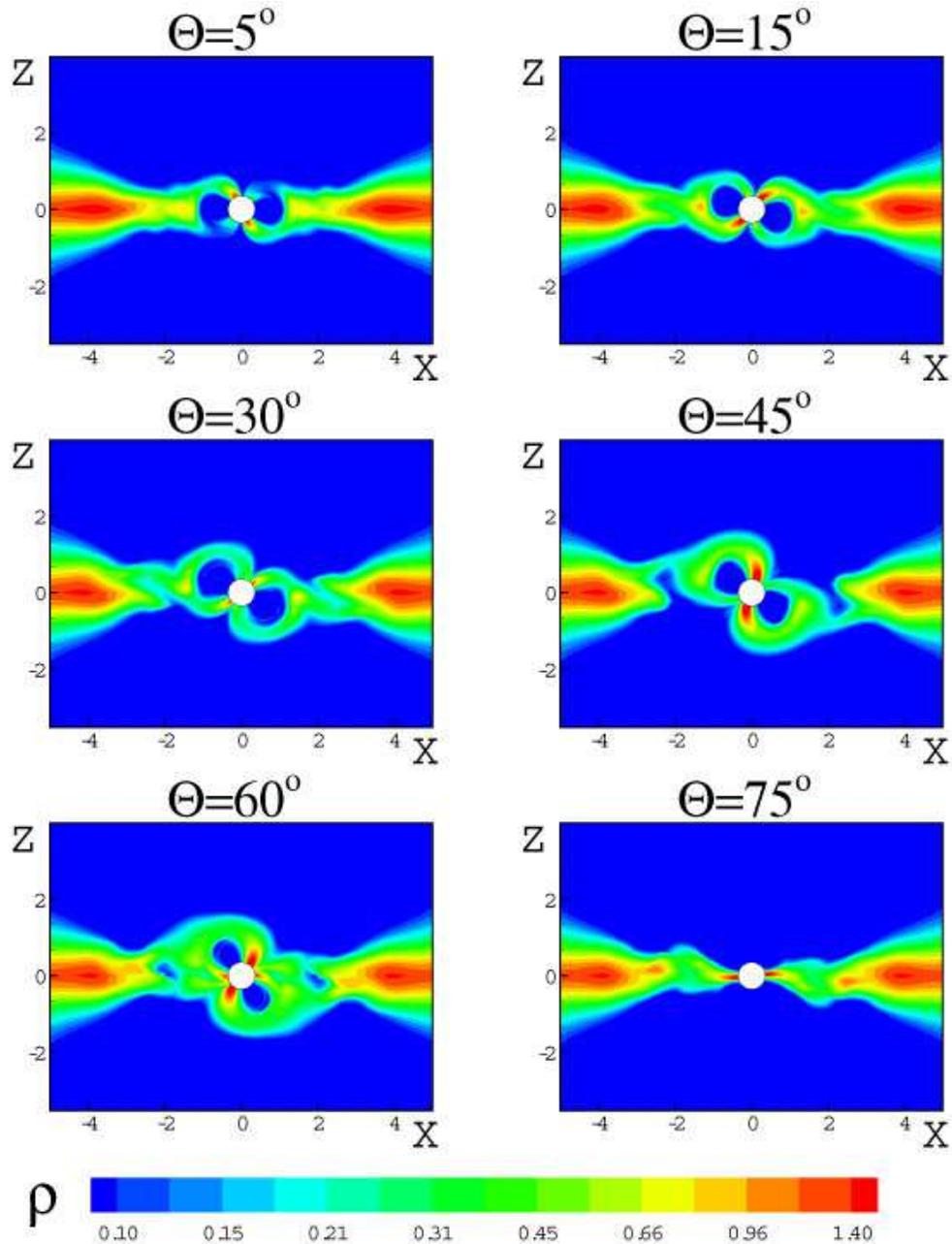} \caption{ The $(X,Z)$ plane
cross-section (in which magnetic moment is located) showing the
density distribution at $T=5$ for different inclination angles.}
\label{Figure 16}
\end{figure*}
%%%%%%%%%%%%%%%%%%%%%%%%%%%%%%%%%%%%%%%%%%%%%%%%%%%%%%%%

%%%%%%%%%%%%%%%%%%%%%%%%%%%%%%%%%%%%%%%%%%%%%%%%%%%%%%%%
\begin{figure*}[t]
\epsscale{1.5} \plotone{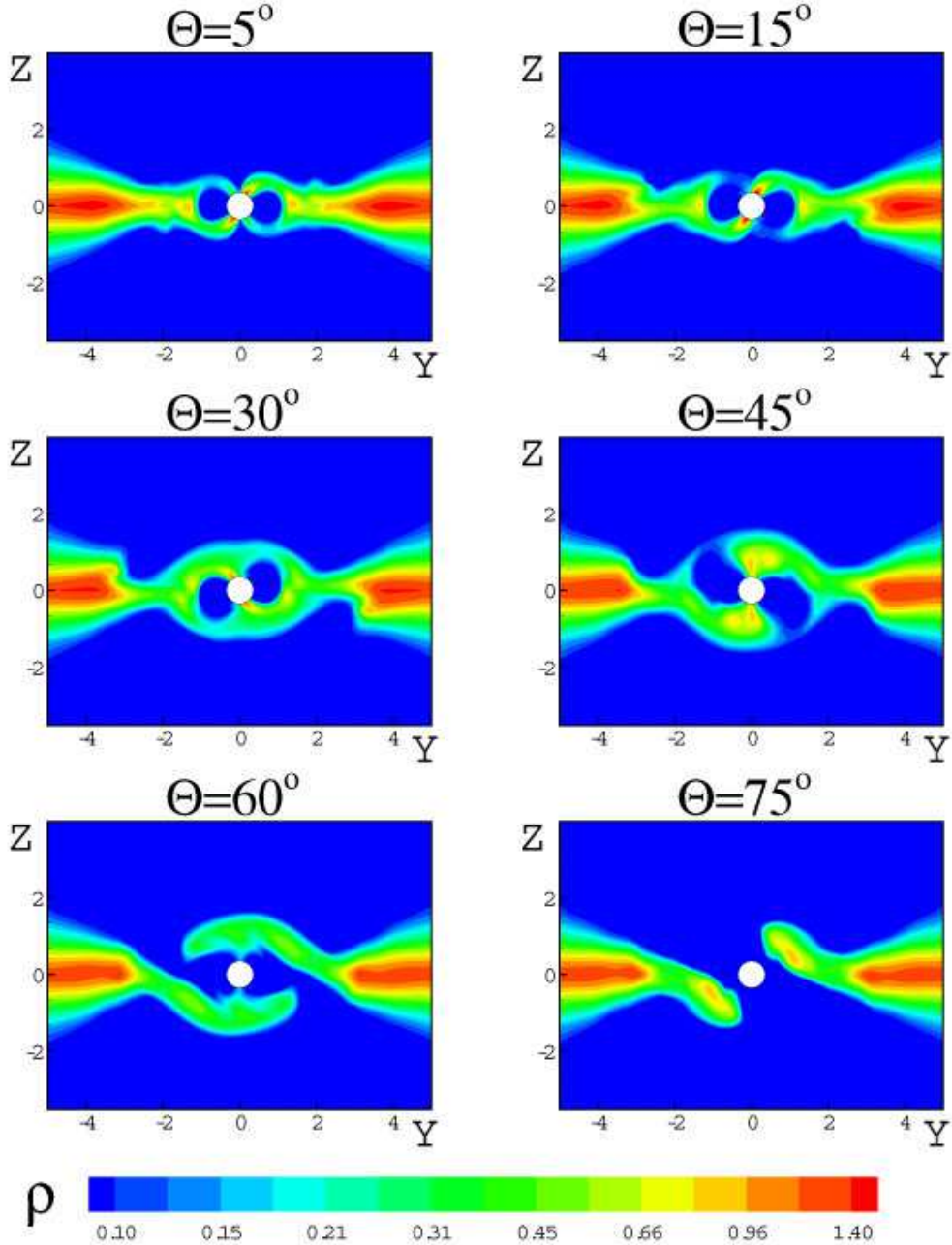} \caption{The $(Y,Z)$ plane
cross-section showing the density distribution for $T=5$ at
different inclination angles.} \label{Figure 17}
\end{figure*}
%%%%%%%%%%%%%%%%%%%%%%%%%%%%%%%%%%%%%%%%%%%%%%%%%%%%%%%%

%%%%%%%%%%%%%%%%%%%%%%%%%%%%%%%%%%%%%%%%%%%%%%%%%%%%%%%%
\begin{figure*}[t]
\epsscale{1.5} \plotone{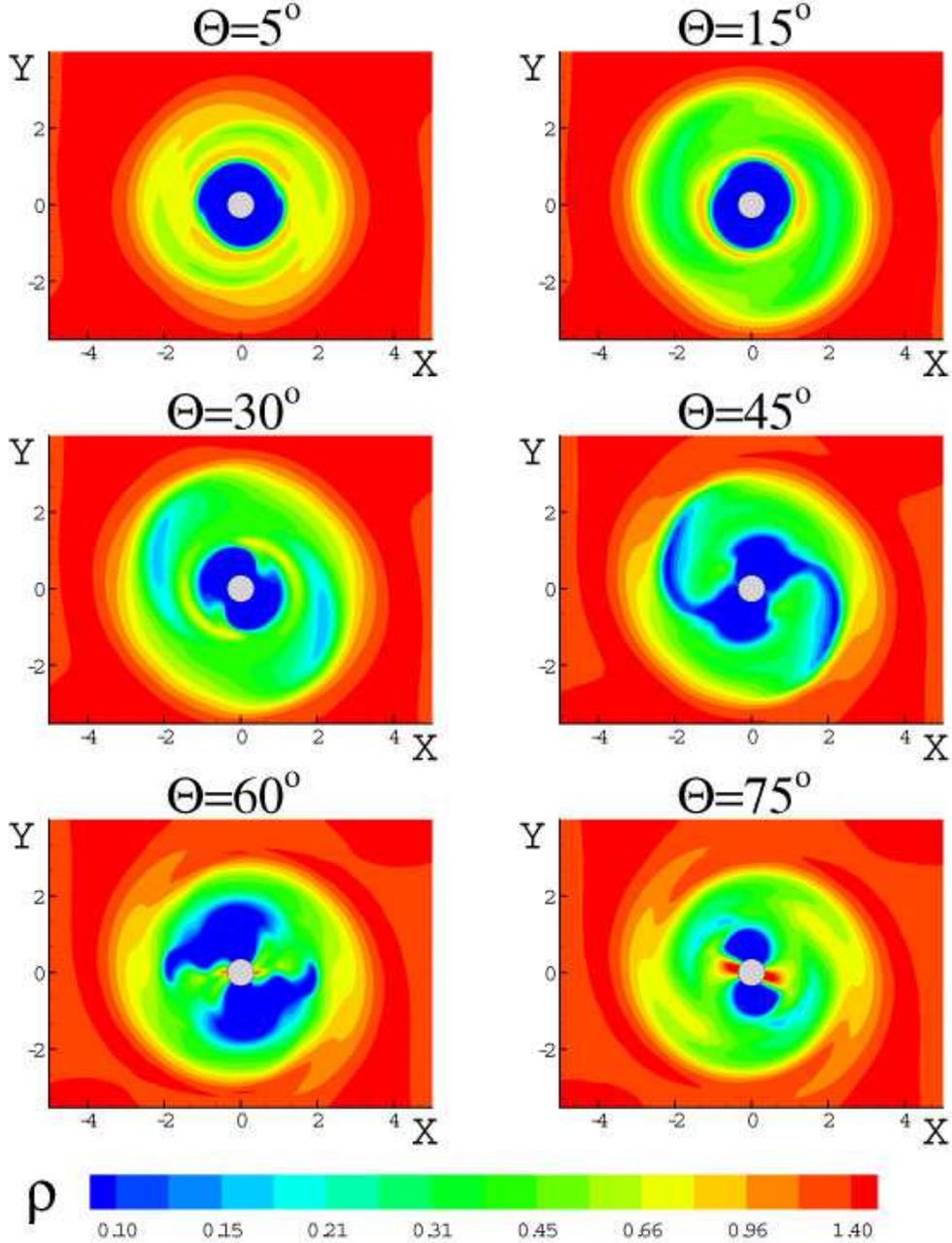} \caption{The $(X,Y)$ plane
cross-section showing the density distribution at $T=5$ for
different inclination angles.} \label{Figure 18}
\end{figure*}
%%%%%%%%%%%%%%%%%%%%%%%%%%%%%%%%%%%%%%%%%%%%%%%%%%%%%%%%

\section{Results of Simulations at Different Inclination
    Angles $\Theta$}

We performed  a number of simulations  for different inclination
angles, $\Theta=2^\circ$, $\Theta=5^\circ$, $\Theta=15^\circ$,
$\Theta=30^\circ$, $\Theta=45^\circ$, $\Theta=60^\circ$, and
$\Theta=75^\circ$. We separately discuss results for small, medium
and large inclination angles.

\subsection{Small Inclination Angles:  $\Theta=2^\circ$, $\Theta=5^\circ$
    and $\Theta=15^\circ$}

First, it is important to investigate the difference between
matter flow at zero, $\Theta=0^\circ$, and {\it very small}
inclination angles. We observed that magnetospheric flow becomes
axisymmetric starting from very small inclination angles,
$\Theta=2^\circ$ and $\Theta=5^\circ$. At such angles, at low
density the ``windows" of even lower density form in the
magnetospheric flow, while at larger densities, matter accretes in
two streams.

       Figure 4 shows initial state (left panel) and magnetospheric
       flow at $T=5$ (right panel) for relatively low density $\rho=0.15$.
       One can see that compared to the case with $\Theta=0^\circ$ (Figure
       2), the windows of lower density form. Windows precess about
       the rotation axis.
        The plots are shown
in the coordinate system rotating with the star,  and  the
$\rvecmu$ vector of the dipole is fixed in the $(X,Z)$ plane and
has the same direction in each of the two plots of Figure 4.
       Precession of
the windows means that its rotation is different from that of the
star.
       Both the star and the disk
rotate counterclockwise looking from the $+ z$ pole towards the
disk.

        The windows rotate in the same
direction but faster which means that the  period of precession is
shorter than that of the star.
      The faster rotation is due to the
fact that inner regions of the disk near the magnetosphere rotate
with an angular velocity close to Keplerian $\omega\approx
\Omega_K(r = 1)\approx 1$ while the star rotates slowly with
$\Omega_*\approx 0.19$.
    The windows  rotate with an angular velocity
    between these two values.

      There is of course a range of densities
in the magnetospheric flow.
      At higher densities, $\rho \gtrsim 0.15$, the windows become larger, and
at even higher densities, $\rho \gtrsim 0.3-0.4$ matter flows in
dense narrow  streams.
       These streams are expected to
be important in the occultation of stellar light from an observer.
       They also
form hot spots where they hit the stellar surface.

The area at the star covered  by the  hot spots depends on
density. At $\rho\approx 0.15$, hot spots cover $A_{spots}\approx
40\%$ of the surface of the star, while at $\rho\approx 0.3-0.4$,
they cover $A_{spots}\approx 20-30\%$ of the surface. At larger
densities of the stream, the area is even smaller (see detailed
analysis in R03).

       Subsequently, we  show  the high density stream component,
but point out that flow pattern  is different at the different
density levels.

       Figure 5 shows the accretion
flow  at the density level $\rho=0.35$. One can see that after
$T=2$ rotations two streams form.
      The azimuthal width of the streams is several times larger
than their thickness in the poloidal direction.
    The streams precess around the $Z$ axis
in the same way as the above mentioned windows.
       For $T \gtrsim 5$, the precession slows, and
the streams settle at a specific location
    about $30^\circ$ downstream of (counter-c
    lockwise from)  the
$({\bf \Omega},\rvecmu)$ plane.

       Earlier, we noticed that the
streams may follow the longer path along the dipole magnetic field
to the star (K02).
       Now we understand that
this is one of the stages of the precession around the $z-$ axis.
Precession slows down because inner regions of the disk come to
co-rotation with the magnetosphere (see also \S 5).
       The precession may
start again if new matter  comes to the inner regions of the disk
with higher  angular momentum.
       We suggest
that if the accretion rate is variable, then this will lead to
higher angular velocity of precession (at larger $\dot M$) and
smaller angular velocity (at lower $\dot M$).

       Figure 6 shows results of simulations
for a  larger inclination angle, $\Theta=15^\circ$.
      One can see that matter
again flows to the star along  two streams.
      However, the streams form more rapidly compared to the
$\Theta=5^\circ$ case, in $T\approx 1.5$ rotations.
      They precess
around the $z-$ axis during $T=1.5-4$.
       Later,
they  settle into a stationary configuration and impact the star
at a location downstream or counter-clockwise from the
$({\bf\Omega,\rvecmu})$ plane.
       Figure 7
shows four projections  at  $T=6$.
     Projection (c) shows that
the streams settled at approximately $30^\circ$ downstream of  the
$({\bf \Omega,\rvecmu })$ plane, the same as in the
$\Theta=5^\circ$ case.

%%%%%%%%new %%%%%%%%
Analysis of forces along the streams has shown that the main
lifting force drugging matter up to the streams is the pressure
gradient force, while the gravity force is the main one dragging
matter along the stream to the star (see more detailed analysis in
R03). The magnetic force is negligibly small. Similar forces were
observed in analysis of funnel flows in axisymmetric  simulations
(R02). However, the inclination of the dipole is favorable for
lifting: the lifting force can be smaller than in the axisymmetric
case.

      The favored locations of the
streams on the magnetosphere can be understood as follows.
      Matter in the inner regions of
the disk spirals gradually inward along almost circular orbits.
      When it comes closer to the
magnetosphere, it is subject to magnetic braking.
       The magnetic braking
is the smallest in the $(X,Z)$ [or, $({\bf \Omega,\rvecmu})$]
plane, where the dipole field lines extend to  the highest and
lowest distances above and below the disk.
     In the $(Y,Z)$ plane,  the
magnetic braking is the largest, because the field lines
symmetrically cross the disk at this plane.
      Thus, matter flows with low braking through
the $(X,Z)$ plane, but is slowed down in the $(Y,Z)$ plane.
  From the other side, the poles are closer to the disk matter in the
$(X,Z)$ plane, and magnetic field lines are directed to the poles,
so that only small lifting force would be sufficient for the
direct accretion to the pole.   In the $(Y,Z)$ plane, matter has
much longer path to the poles, and larger lifting force should be
applied for accretion in the $(Y,Z)$ plane.
       Finally, it seems that matter ``prefers" to
flow in between of $(X,Z)$ and $(Y,Z)$ planes.
       It is
important  to note that for the considered slowly rotating star,
the inner part of the  disk  rotates faster than the
magnetosphere.
     Thus the stream settles
     downstream of (counter-clockwise from) the
     $({\bf \Omega},\rvecmu)$ plane.
      Note, that the
location of the stream is different for different values of
$\Omega_*$ (see R03).

Figure 6  shows that for $T
   \gtrsim 4$, magnetic braking leads to a
partial disruption of the disk at $r < r_{br}\approx 2-3$.
      In this region of the disk, the density
becomes lower with time.
      From the other side, matter accumulates near the
magnetosphere boundary ($r_m\approx 1-1.5$) forming a ``ring",
where the density of the disk increased.  Matter from the ring
gradually accretes to the star.
      This matter
distribution is similar to that observed in two-dimensional
simulations of R02. The difference is that in 3D simulations
matter  often forms a spiral structure in the area of the magnetic
braking $r < r_{br}$.

Figures 6 and 7 show that the inner region of the disk becomes
asymmetric and the projection (d) shows that inner parts of the
disk are {\it ``lifted''} above the $Z=0$ plane of the distant
disk.
       Observations of different
cases show that this warp is connected with the tendency of the
accreting matter to co-rotate with the magnetosphere.
       The inner region
of the disk is tilted so as to have its normal vector in the
direction of the magnetic moment $\rvecmu$. This direction of the
warp is different from that predicted theoretically (see \S 1).

Simulations at small inclination angles have shown  that
    {\it even a small inclination
of the dipole moment} leads to non-axisymmetric accretion and
possible variability of the light emitted from the surface of the
star.

\subsection{Medium Inclination
Angles: $\Theta=30^\circ$ and $\Theta=45^\circ$}

       Figure 8 shows the matter flow for an inclination angle
$\Theta=30^\circ$.
      We observed that the dense funnel flow streams form
faster, after about one rotation.
        At first, two prominent streams
form and precess as in the case of small inclination angles.
      However, at $T=3-4$ the
    magnetic braking starts
to disrupt the inner regions of the disk ($r \lesssim 2-3$), and
the streams turn into leading spirals. We do not bring new matter
with Keplerian angular momentum, so that the density in the inner
regions of the disk  gradually decreases and matter starts to
rotate with sub-Keplerian velocity.
      At the same time, a dense
ring  forms around the magnetosphere of the star as in the cases
of smaller $\Theta$.
      Later, at $T \gtrsim 6$, the density
decreased further and a wide gap formed between the inner disk and
the star.
       There
is still low density matter between the inner ring and the inner
regions of the disk.
      This structure is similar
to that observed in two-dimensional simulations: an inner ring
forms around the magnetosphere, and there is a lower density disk
around it.

      Figure 9 shows different projections at $T=6$.
Projection (d) shows that the inner region of the disk is warped.
      The warp arises from the tendency of the accreting matter to
corotate with magnetosphere; that is, the normal to the warped
part of the disk tends to align with $\rvecmu$.

Figure 10 shows the accretion flow at $\Theta=45^\circ$.
      At this relatively large inclination angle,
the accreting matter forms a complicated structure with initially
two streams as in other cases.
      But later, four streams form.
       Inner regions of the disk
experience  magnetic braking and the disk is significantly
disrupted at the chosen density level.
       However, there is always a bridge of
matter connecting external regions of the disk with magnetospheric
flow.
      This bridge is connected with
the fact that some matter accretes to the poles ``directly" as a
result of significant inclination of the dipole.
     Such multiple streams may give rise to
complicated variability of emissions from the stellar surface.

Two sets of magnetic field lines are shown in Figure 10.
        One set
is the same as in Figures 5, 6 and 8. They thread the disk at
$r\approx 2.8$ and are closed inside the region $r < 3$.
       Another set of field
lines start closer to the poles.
      Some of these
lines thread the inner regions of the disk due to high inclination
of the dipole; others  thread the disk far away, $r>>3$. The field
lines threading the disk at close distances, become involved in
the disk rotation  and form azimuthal magnetic field inside the
disk.

      Figure 11 shows different
projections of the flow for $\Theta=45^\circ$ at time $T=7$. The
projection (a) shows that the flow is symmetric if the observer
looks along the $\rvecmu$ axis.
      Projection (c) shows that the streams
associated with ``direct" accretion to the poles are located at
the angle $\sim 20^\circ - 30^\circ$ downstream (counter-
clockwise) from the $ ({\bf \Omega},\rvecmu)$ plane as in all
above considered cases with smaller $\Theta$.
      Projection (d)
shows that magnetospheric flow has a tendency to co-rotate with
magnetosphere as found earlier for smaller $\Theta$'s. The funnel
flows rise high above the disk plane.

\subsection{Large Inclination Angles:
$\Theta=60^\circ$ and $\Theta=75^\circ$}

Figure 12 shows the evolution of the matter flow for
$\Theta=60^\circ$.
      The flow becomes complicated  after $T \gtrsim 1$.
      From one side,
matter accretes directly along field lines to the pole which (at
such inclination) is close to the plane of the disk.
      From the other side, the matter goes
around the  magnetosphere and accretes through the remote path
along the dipole field lines.
      The combination of these two flows
gives a complicated overall flow.
       At $T=7, 8$, the direct
polar flow predominates.
      However, at $T=9$, more matter came from
the disk and matter flow becomes more complicated.

       Two sets of magnetic field lines
are shown in Figure 12.
      The number of ``polar" field lines
threading the inner disk
    is larger than that at $\Theta=45^\circ$
due to the dipole inclination.
      The magnetic field lines
are dragged by the disk forming azimuthal field inside the disk.

Figure 13 shows four projections of the $\Theta=60^\circ$ flow at
$T=8$.
      One can see that the flow
looks very different at different projections.
      Projection (c)
shows that most of matter accretes through the nearby pole of the
inclined dipole.
      The location of the streams is slightly downstream of
the $({\bf \Omega},\rvecmu)$ plane.
      Complicated warping is observed which is
connected with magnetospheric accretion.

Lastly we discuss the high inclination angle flow with
    $\Theta=75^\circ$.
In this case many  ``polar" field lines thread the inner region of
the disk (see Figure 14 at $T=0$), while the closed field lines
cross the path of the disk matter.
         In such a configuration,
the disk matter experiences strong magnetic braking, and it
    accretes to the poles
starting from $T < 1$. Some matter, however, tries to continue its
``circular" path around the star. This matter is lifted up and
goes around the magnetosphere, because the closed magnetosphere is
on the way.
    The magnetic field lines which initially were
closed inside $r < 3$, moved closer to the star dragged by
accreting matter as for other $\Theta$'s. The ``polar" magnetic
field lines show complicated bending. Some of them are dragged by
the disk, forming an azimuthal field inside the disk.

Figure 15 shows projections of the matter flow at $T=5$.
      Projections
(a) and (d) show that matter flowing around the magnetosphere is
strongly lifted above the disk plane.
       These features of the matter flow
    are typical for a slowly rotating star,
where matter in the inner regions of the disk  rotates at a higher
angular speed than the star and its magnetosphere.
        Projection (c)  shows that the matter
streams reach the star downstream of the $({\bf \Omega},\rvecmu)$
plane as seen in
    other cases.
      We conclude that this  position of the
streams at the magnetosphere
%at the star
is a general feature during accretion to a
slowly rotating star (see also  panels (c) in Figures 7, 9, 11,
13, and 15).
      Note that in  cases of
a faster rotating star (closer to equilibrium) or in the case of a
very fast rotating star (``propeller" regime), the location of the
streams and their shapes are different (see R03).

\subsection{Warping of the Disk}

  In all of our simulations we
observed that the inner regions of the disk are disturbed and
become non-axisymmetric.
    Some parts of the disk are above and
others are below the equatorial plane so that the inner disk is
warped.
    In all cases we observed that this warp reflects the
tendency of matter to co-rotate with magnetosphere of the star, so
that the normal of the inner warped disk has tendency to coincide
with the dipole moment $\rvecmu$.
   This direction of warping does
not coincide with any of the proposed theories.
    Such a warp is
specifically noticeable at the Figure 9d, for $\Theta=30^\circ$.
     This warp has a different nature.  It
represents the matter flow around the magnetosphere and is not a
part  of the disk.   One can see that matter is warped in the
direction of the remote pole.

    A possible reason that the observed warp does not coincide
with the theory is that most of theories deal with pure
diamagnetic disks (e.g., Aly 1980; Lipunov \& Shakura 1980). Lai
(1999) considered non-diamagnetic and partially diamagnetic disk,
but his lifting and precessional forces mainly rely on the
twisting of the magnetic field lines, that is on formation of
strong $B_\phi$ component above the disk.
    We did not observe
significant twisting of the field lines in the inner regions of
the disk or in the magnetosphere.
    In the regions of the disk very close
to the magnetosphere,   $r_m \lesssim r \lesssim 2-3$, the disk
has tendency to corotate with the star.   Also, the corona just
above the disk has a tendency to corotate with the disk (see also
detailed analysis in R02).  For this reason the twisting of the
magnetic field is small.
     At  larger distances, $r \gtrsim 3$,  the azimuthal component
of the magnetic field may be comparable to the
   poloidal component.
However, these field lines have tendency to be open.
   Further, the magnetic field in this region is not strong enough to
influence the dynamics of the disk.
     We conclude that the
magnetic field lines
   threading the inner regions of the disk are not twisted sufficiently
to give  this type of warping. Further analysis is needed for
understanding of warping and asymmetries observed in simulations.
Special simulations with  a thin diamagnetic disk may be used to
check the theory.

\section{Analysis of the Magnetic Braking}

      The simulations show that the interaction
of the star's magnetic field with the inner regions of the disk
causes a redistribution of density in this part of the disk.
      The three-dimensional pictures presented
above show the magnetospheric flow at fixed density levels.
      To show the spatial distribution of density,
we use two-dimensional slices.
     We also show
one-dimensional distribution of density and angular momentum
density along different axes.

     To compare results obtained
at different inclination angles $\Theta$, we take a fixed
simulation time.
       We were able to run some
cases longer than others, reaching up to $10-12$ rotations in some
cases.
      However, in order to incorporate the
case  with the highest inclination, $\Theta=75^\circ$, we chose
time $T=5$ as the reference time for comparing results at
different $\Theta$.

        Figures 16-18 show  two-dimensional
slices of the density distribution for different inclination
angles.
     Figure 16 shows the slice of the
$(X,Z)$ plane (at $Y=0$), which coincides with the plane
containing the  magnetic moment (the $({\bf \Omega},\rvecmu)$
plane).
      From Figures 16 and 17 it is seen
that the external regions of the disk, at $r \gtrsim 3-4$, are not
appreciably disturbed and are approximately the same for different
inclination angles $\Theta$.
      Inside this region, however,
the scales at which the
    magnetosphere influences
the matter flow are different.
      Figure 16 shows that for
$\Theta=5^\circ$, matter starts to go into funnel
    flows around the magnetosphere at
approximately  $r\approx 1.5$.
      For $\Theta=15^\circ$, this occurs
at $r\approx 1.8$; at $\Theta=30^\circ$,
     at $r\approx 1.9$. At $\Theta=45^\circ$,
$60^\circ$, and $75^\circ$, the region is even larger, $r\approx
2.2$.
       Figure 17 shows the density distribution in
the $(Y,Z)$ plane (at $X=0$).
       This cross-section also shows that at
larger $\Theta$, the magnetospheres are larger.
      Figure 18 shows the
$(X,Y)$ cross-sections ($Z=0$), which go through the plane of the
disk.
      Figure 18
shows that a relatively high density ring (white color) forms
around the low-density magnetospheres at small inclination angles,
$\Theta=5^\circ$, $15^\circ$ and  $30^\circ$, but it disappears at
larger inclination angles.
      At very large
inclination angles, $\Theta=60^\circ$ and  $75^\circ$, the direct
accretion to the poles is observed (white color in the polar
regions).
             This cross-section also shows that the inner
region with radii $R \lesssim 3-4$, has a lower density than the
rest of the disk.

%%%%%%%%%%%%%%%%%%%%%%%%%%%%%%%%%%%%%%%%%%%%%%%%%%%%%%%%%%
\begin{figure*}[t]
\epsscale{1.4} \plotone{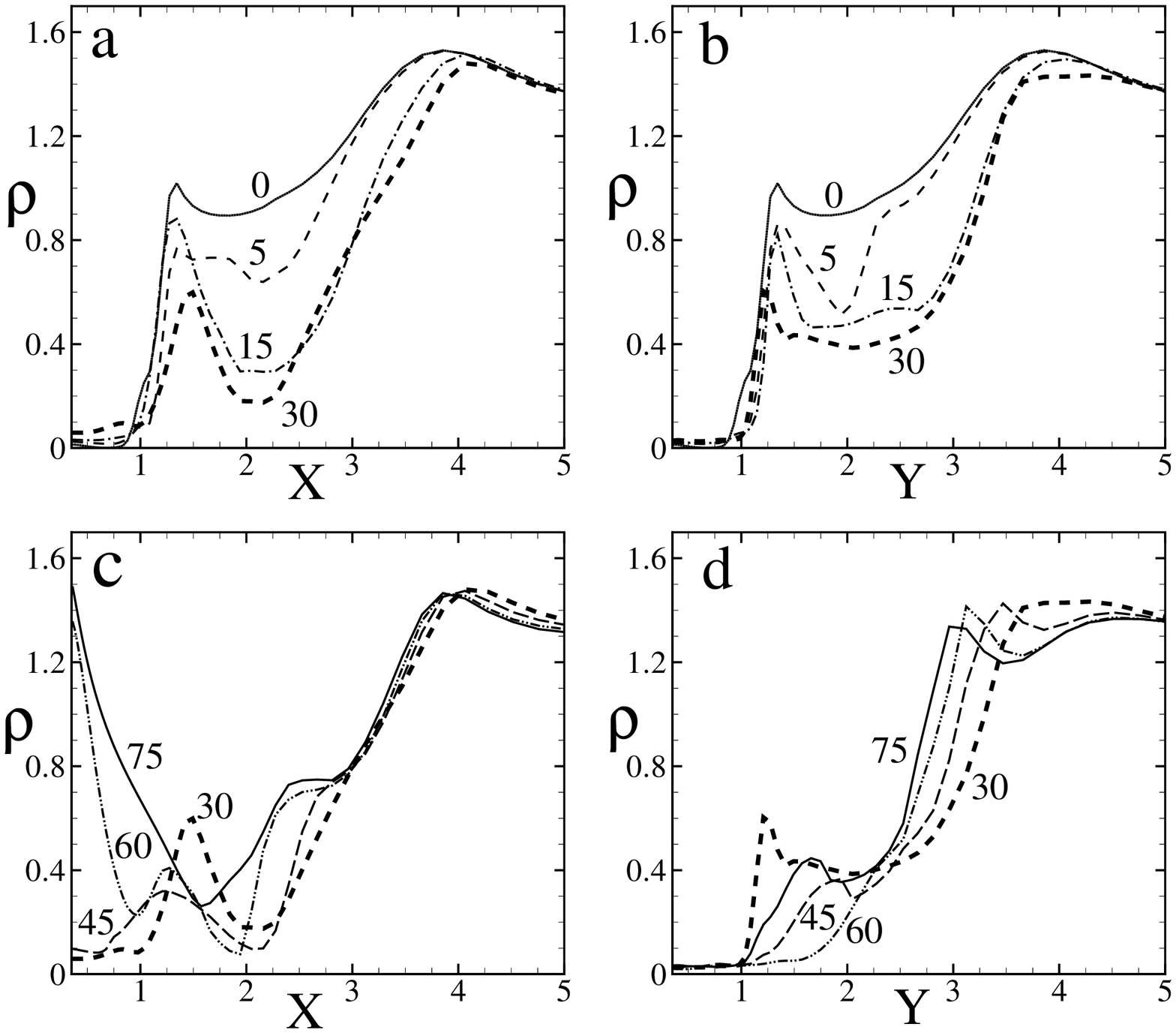}
    \caption{Density distribution along the $x$ (left panels)
    and $y$ (right panels) axes for
    inclination angles $\Theta=0^\circ,~ 5^\circ,
~ 15^\circ,~ 30^\circ$ (top panels)
    and $\Theta=30^\circ,~ 45^\circ, ~60^\circ,
~ \& 75^\circ$ (bottom panels).} \label{Figure 19}
\end{figure*}
%%%%%%%%%%%%%%%%%%%%%%%%%%%%%%%%%%%%%%%%%%%%%%%%%%%%%%%%%%

We also did {\it one-dimensional} analysis of the density and
angular momentum distribution along the $X$ and $Y$ axes.
     For clarity of the
presentation, we divided the inclination angles into two groups:
smaller angles, $\Theta= 0^\circ$, $5^\circ$, $15^\circ$, and
$30^\circ$, and larger angles: $\Theta= 45^\circ$, $60^\circ$,
$75^\circ$.
      The case $\Theta=
30^\circ$, was added to  the second group  for convenience of
comparisons.
      Figure 19a shows density distribution along the
$X$ axis for smaller inclination angles.
       One can see that the disk
is disturbed by magnetic braking for $r \lesssim 4$.
       One can see that at the larger
    inclination angle the density is smaller.
       The density is
    very small for $X \lesssim 1$, where the
magnetic field is very strong.
      The strong magnetic field limits the matter flow.
      At larger $X$, the density increases forming
a peak of density.
         This is the ring mentioned earlier, where matter accumulates
before accreting to the poles.
       This feature was the typical
one in all two-dimensional simulations (see R02).
        It is  also clearly
observed in the simulations at  small inclination angles.
        Figure 19b shows similar distributions
along the $Y$ axis, which are similar to the distributions along
the $X$ axis.

      At larger inclination angles, $\Theta=45^\circ, 60^\circ$ and
$75^\circ$,  the density distribution is different (see Figures
19c,d).
       In the $X$ direction,
for $\Theta=45^\circ$, matter comes closer to the stellar surface.
      For $\Theta=60^\circ$ and $\Theta=75^\circ$, the density
strongly increases towards the star, which means that matter
accretes directly to the star.
       The magnetoshere is not an obstacle for the
flow at  high inclination angles $\Theta$.
       In the $Y$ direction (see Figure 19d),
the matter is stopped by the magnetospheric at a radius $r\approx
1$ for all inclination angles.

%%%%%%%%%%%%%%%%%%%%%%%%%%%%%%%%%%%%%%%%%%%%%%%%%%%%%%%%%%
\begin{figure*}[t]
\epsscale{1.4} \plotone{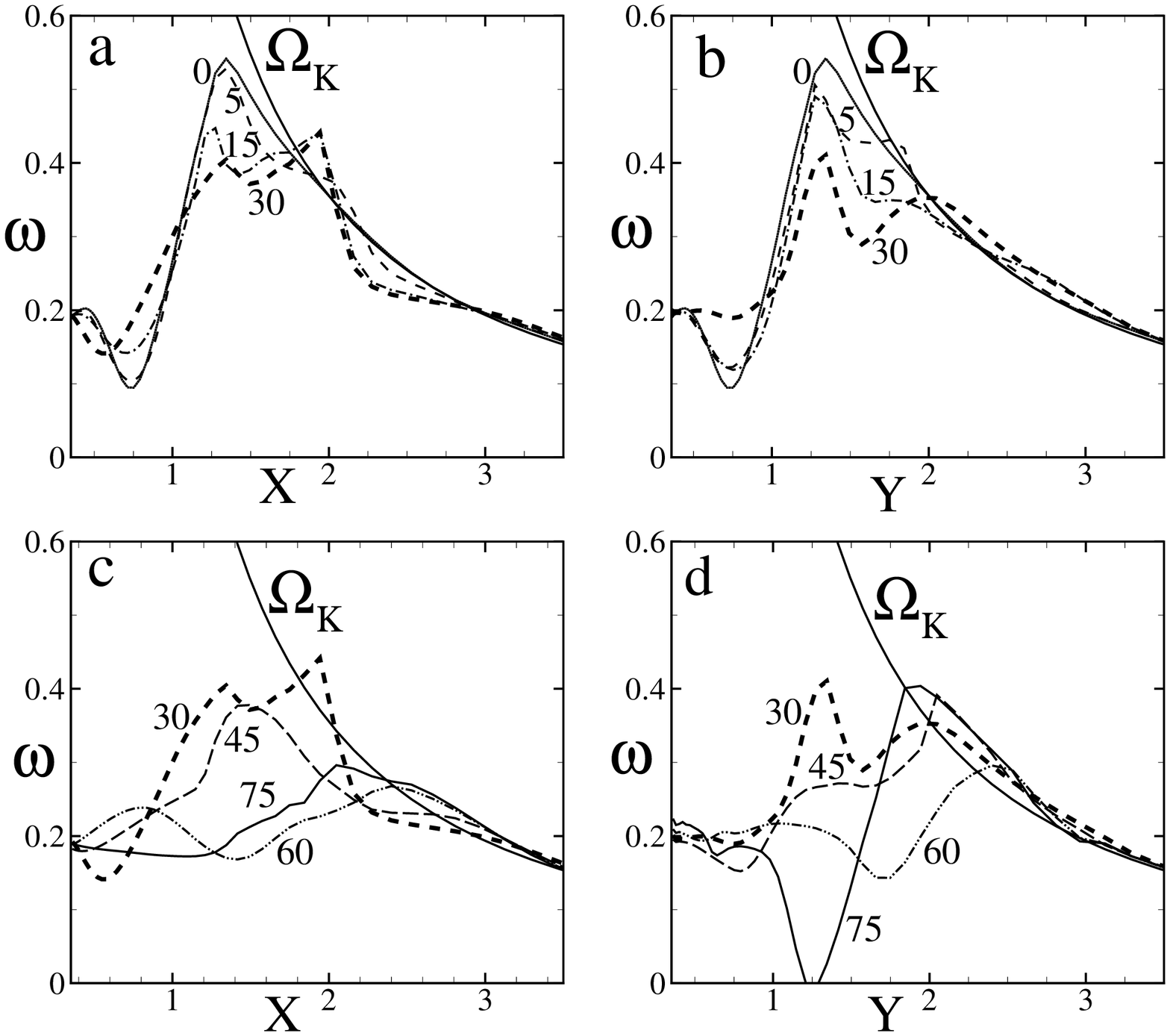}
    \caption{Angular velocity distribution
along the $X$ axis (left panels)
    and $Y$ axis (right panels)  for
    inclination angles $\Theta=0^\circ,~
    5^\circ,~ 15^\circ,~ 30^\circ$ (top
panels)
    and $\Theta=30^\circ,~ 45^\circ,~
60^\circ,~ \& ~75^\circ$ (bottom panels).} \label{Figure 20}
\end{figure*}
%%%%%%%%%%%%%%%%%%%%%%%%%%%%%%%%%%%%%%%%%%%%%%%%%%%%%%%%%%

%%%%%%%%%%%%%%%%%%%%%%%%%%%%%%%%%%%%%%%%%%%%%%%%%%%%%%%%%%
\begin{figure*}[t]
\epsscale{1.4} \plotone{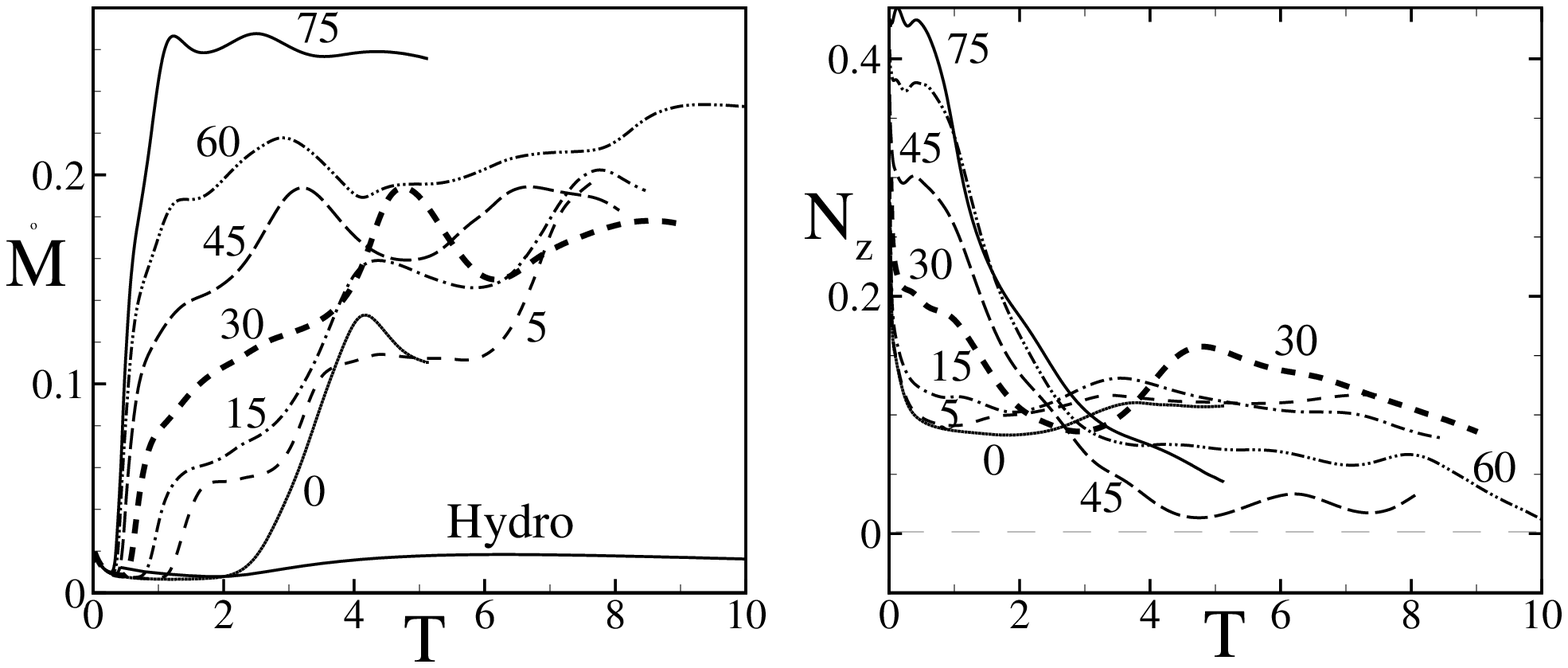}
    \caption{The left-hand panel shows the
    matter flux to the star $\dot M$ versus
time $T$ for different inclination angles $\Theta$.
       The right-hand panel shows
torque $N_z$ versus time at different $\Theta$. } \label{Figure
21}
\end{figure*}
%%%%%%%%%%%%%%%%%%%%%%%%%%%%%%%%%%%%%%%%%%%%%%%%%%%%%%%%%%

       Magnetic braking leads to the deviation of the angular velocity
of the matter from the Keplerian value.
      Figures 20a, b show that at radii $r \lesssim 3$,
the angular velocity $\omega$ deviates from Keplerian value
starting from $r \lesssim 3$.
       It gradually become smaller than
Keplerian and at $r \lesssim 1.3$ it sharply decreases  to the
angular velocity of the star $\Omega_*=0.19$.
       At large inclination angles (see Figures 20c,d) the
deviation from Keplerian rotation is much stronger in both the $X$
and $Y$ directions.
      For $\Theta=60^\circ$ and
$75^\circ$, the angular velocity inside the region $r \lesssim 3$
is $\omega\approx 0.2-0.3$; that is,
    the inner regions of the disk almost co-rotate with the star.

\section{Matter and Angular Momentum Fluxes}

Here, we analyze the mass accretion rate to the surface of the
star for different inclination angles.
      Also, we analyze the angular momentum flux to the
star and discuss its influence on the angular momentum of the
star.

\subsection{Matter Flux}

We calculated the matter flux to the surface of the star,
    $$
\dot M=-\int d{\bf S}\cdot{ {\rho{\bf v}}}~,
    \eqno(5)
$$
where the surface integral is taken just outside of the star with
$d{\bf S}$ the outward pointing surface area element.

      Figure 21 (left panel) shows the
matter fluxes at different inclination angles versus time.
       Initially the accretion rate is zero,
but it increases rapidly as matter reaches the surface of the
star.
       It takes less time for larger
inclination angles.
      After the initial rise,
the accretion rate settles at a value which changes only slowly,
forming the ``plateau".
       At $\Theta=30^\circ$ and
$45^\circ$, the plateau values of $\dot M$ are approximately the
same, $\dot M\approx 0.19$.
      At $\Theta=60^\circ$,
$\dot M\approx 0.2-0.24$, and at $\Theta=75^\circ$ it is even
larger, $\dot M\approx 0.26$.
    At the smallest inclination angles, $\Theta=5^\circ$ and
$\Theta=15^\circ$, the $\dot M$ increases gradually and reaches
the values $\dot M \approx 0.19-0.2$, but it is not clear, whether
the plateau was formed or not.
      Note, that  at $\Theta=0^\circ$,the
accretion rate is smaller, $\dot M\approx 0.13$, but not
significantly smaller than that for inclined rotators.
       These
values of $\dot M$ are in accord with two-dimensional axisymmetric
simulations done with a similar grid, $\dot M\approx 0.14-0.16$.
       Note, that simulations in the hydrodynamic
case ($B=0$) lead to a much
    smaller accretion rate $\dot M\approx 0.02$.
       This accretion is due to the
small numerical viscosity as discussed in the Section 2.
       Thus, we conclude that the  accretion rate is larger at larger
$\Theta$ because the magnetic braking is larger at larger
$\Theta$.

%%%%%%%%%%%%%%%%%%%%%%%%%%%%%%%%%%%%%%%%%%%%%%%%%%%%%%%%%%%%%
\begin{figure*}[t]
\epsscale{1.4} \plotone{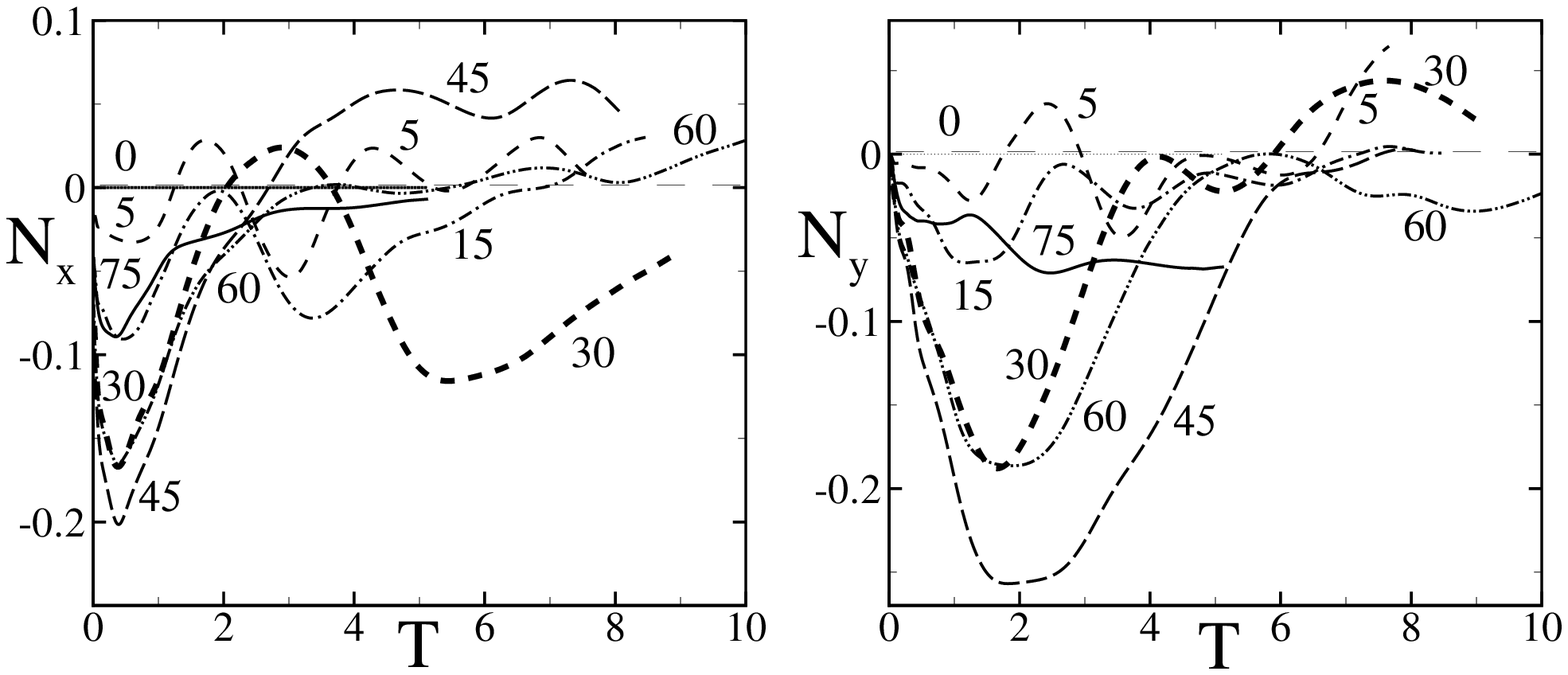} \caption{Torque on the star
along the $X$ axis (left panel) and $Y$ axis (right panel) at
different inclination angles $\Theta$.} \label{Figure 22}
\end{figure*}
%%%%%%%%%%%%%%%%%%%%%%%%%%%%%%%%%%%%%%%%%%%%%%%%%%%%%%%%%%%%

\subsection{Torque on the Star}

       Incoming matter and the
magnetic field carry in or remove angular momentum from the star.
The angular momentum of this matter is much less than that of the
star  during the time span of the simulations. Thus, we fix angle
$\Theta$ and  direction and value of $\bf\Omega$. We calculate
torques relative to different axes which carry information about
possible long-term evolution of the star.

       The total angular momentum flux
in the reference frame rotating with the star $(X,Y,Z)$ consists
of the part carried by the matter which gives the matter torque,
${\bf N}_m$, and that carried by the field which gives the field
torque ${\bf  N}_f$,
$$
     {\bf  N}= {\bf  N}_m + {\bf  N}_f
= - \int d{\bf S}\cdot {\bf v}~ \rho~ ({\bf r \times u})
     + \frac{1}{4 \pi} \int d{\bf
S}\cdot {\bf B}~
     ({\bf r \times  B}) ,
\eqno(6)
$$
  where the integration is over a
spherical surface just outside of the star with $d{\bf S}$ the
outward pointing surface area element, and where ${\bf u}$ is the
velocity in the inertial frame and ${\bf v}$ is the velocity in
the frame rotating with the star.
       For an inclined rotator,
${\bf N}$ is not in general along a particular axis.
       Denoting ${\bf L}$ the angular
momentum of the star, we have
$$
{d {\bf L} \over dt} = {d^\star {\bf L} \over dt} +{\bf \Omega
\times} {\bf L} = {\bf N}~ \eqno(7)
$$
where $d/dt$ is with respect to an inertial reference frame and
$d^\star /dt$ is with respect to the rotating $(X,Y,Z)$ coordinate
system.
    For the considered case,  the
angular momentum of the star and that of the incoming matter in
the disk have the same direction.
   Therefore,
the direction of ${\bf L}$ averaged over the star's rotation
period remains in the $z$ direction of the inertial or laboratory
reference frame.

         We observed that in the
vicinity of the star the angular momentum flux is carried
predominantly by the magnetic field.
       The flux carried by the
     matter is very small,  about
$1\%$ of the total.
        Far  from  the magnetosphere,
the angular momentum flux is carried mainly by the matter.
        With decreasing distance, the
matter transfers its angular momentum to the magnetic field.
        A similar behavior was observed
in   axisymmetric simulations (R02).

        The torque about the $Z$ axis,
$N_z$, is responsible for spinning-up or spinning-down of the
star.
       Positive ${N}_z$ corresponds to
spinning-up of the star.
       Figure 21 (right panel) shows
the flux $N_z$ for different inclination angles.
        In the present work the star
rotates slowly so that the incoming matter spins-up the star.
        The variation of  $N_z$  is
analogous to the variation of $\dot M$ (Figure 21, left panel).
        One can see, that initially,
the magnetic braking was large for all inclination angles, but
later it decreased and became relatively constant.
         At small inclination angles,
$\Theta=0^\circ, 5^\circ$ and $15^\circ$, the torque went to an
approximately constant value rapidly, $T < 0.5$.
         The torque is almost constant
for $\Theta=0^\circ, 5^\circ$ , and decreases only slowly  at
$\Theta=15^\circ$.
        At large $\Theta$, the $N_z$
decreases  gradually.
         If new matter is accreted to
the inner regions of the disk, then $N_z$ will stay at some
non-zero level, as we observed in our two-dimensional simulations
(R02).
        Additional simulations with
steady accretion from the disk are needed to resolve the
difference in angular momentum transport at different $\Theta$.

       A positive torque $N_x$ would,
in the absence of the ${\bf \Omega}\times {\bf L}$  term, act to
shift the star's angular momentum ${\bf L}$ in the $+X$ direction,
that is towards the direction of $\rvecmu$.
        However, the ${\bf
\Omega}\times {\bf L}$ is expected to be important for an actual,
oblate star where the moments of inertia are $I_z >I_x = I_y$ with
$(I_z-I_x)/I_x \ll 1$.
   For this reason a torque $N_x$
acts to give $\Omega_y = N_x/(I_x\Omega_z \Omega_{prec})$ which
corresponds to a very slow precession of the $\bf\Omega$ about
${\bf L}$ which is fixed in space, where $\Omega_{prec}
=(I_z-I_x)\Omega_z/I_x$ is the free precession angular frequency.
    Figure 22 (left panel)
shows the torque $N_x$ for different inclination angles $\Theta$.
    One can
see, that at $\Theta=0$, $N_x=0$, which confirms that our
three-dimensional code supports axisymmetry with high accuracy.
     This also confirms that the
origin of this momentum is connected with interaction of the {\it
inclined} dipole with the disk.
       Even at small inclination
angle, $\Theta=5^\circ$, the torque $N_x$ is non-zero.
       It varies  sinusoidally and
changes sign from negative to positive and back, and settles to a
positive value.
       For $\Theta=30^\circ$,
variations of larger amplitude were observed.
       At  inclination angles,
$\Theta=45^\circ$ and $60^\circ$, $N_x$ is initially negative, but
becomes positive later.
      A negative torque at
$\Theta=75^\circ$ gradually increases to larger values.
%In this
%case the flux is always negative, because matter is always lifted
%above the magnetosphere downstream of the $({\bf \Omega},\rvecmu)$
%plane.

       A positive torque $N_y$ would,
in the absence of the ${\bf \Omega}\times {\bf L}$  term, act to
shift the star's angular momentum ${\bf L}$ in the $+Y$ direction,
that is towards the direction of ${\bf \Omega}\times\rvecmu$.
        However, as mentioned the ${\bf
\Omega}\times {\bf L}$ is expected to be important for an actual,
oblate star.
   For this reason a torque $N_y$
acts to give $\Omega_x = -N_y/(I_x\Omega_z \Omega_{prec})$ which
corresponds to a very slow precession of the $\bf\Omega$ about
${\bf L}$ which is fixed in space.
       Figure 22 (right panel) shows
the torque $N_y$ for different inclination angles.
       We observed that for $\Theta=
0$, $N_y=0$, as it should be.
       At the small inclination angle
$\Theta=5^\circ$,  $N_y$ varies around zero value, which is
connected with the formation of  two streams which precesses
around the magnetosphere of the star.
       At $\Theta=15^\circ$, the sign
of $N_y$ is mainly negative, though variations of the flux are
also observed.
     At $\Theta=30^\circ$ and
$\Theta=60^\circ$, the initial variation of the flux at $T
\lesssim 2$ has the same value,
     $N_y\approx - 0.18$.
           However,  later both fluxes
decreased to smaller values.
     Note that  at $\Theta=30^\circ$,
the $N_y$ changes sign to positive values.
       For $\Theta=75^\circ$, the
torque $N_y$ goes to an approximately constant value $\approx -
0.07$.

        We emphasize that for more
rapidly rotating stars, the torques may behave quite differently
from the case of the slowly rotating star considered here.

\section{Conclusions and Observational Consequences}

We performed full three-dimensional ideal MHD simulations of disk
accretion to a slowly rotating star with an inclined dipole
magnetic field for different inclination angles $\Theta$.
   Below, we summarize the main results
of the simulations (\S 7.1) and discuss possible observational
  consequences   (\S 7.2).

\subsection{Main Results of Numerical Simulations}

     In the following we summarize the main
results and conclusions of our simulations.

{\bf 1.}~       We observed that the accretion flow near the star
becomes non-axisymmetric at very small inclination angles, $\Theta
\sim 2^\circ - 5^\circ$.

{\bf 2.}~ The shape of magnetospheric flow depends on the density.
    At  a sufficiently low
density, most of the volume of the magnetosphere is filled by
matter, excluding small ``windows'' of even lower density.
     At a higher density, the
matter is in two or several ``accretion streams.''

{\bf 3.}~  For relatively  small inclination angles, $\Theta
\lesssim 30^\circ$, the densest matter accretes in {\it two
streams}.
    The streams typically follow a path to
the closest magnetic pole of the star with displacement of
$20^\circ-30^\circ$ downstream (counter-clockwise) from the $({\bf
\Omega},\rvecmu)$ plane on the star's surface.

    The accretion streams may  precess about the
star's rotation axis.
    The precession appears
to be  caused by  incoming matter with
  higher than average angular momentum;
this matter causes the streams to rotate faster than the star.
    In periods of steady
accretion, the streams do not precess.
   Precession is observed
more often in cases with small $\Theta$.

  For intermediate inclination angles,
$30^\circ \lesssim \Theta \lesssim 60^\circ$, the magnetospheric
flow consists of two or more streams.
    Several streams
(typically four) is a common feature for $\Theta=45^\circ$, while
in other cases ($\Theta=30^\circ$, $\Theta=60^\circ$) several
streams appear in periods of enhanced accretion.
   The number of streams
may depend on the  density level.

    For large inclination angles,
$\Theta \gtrsim 60^\circ$, matter typically accretes in two
streams.
  However, the streams
have a different shape compared with those at small $\Theta$
  and they come to
the star near the equatorial plane.

{\bf 4.}~  The   density distribution  is close to that observed
in two-dimensional simulations (R02). Namely, magnetic braking
leads to depression of density in the region $r_m \lesssim r
\lesssim 4$ and to the formation of a dense ring near the
magnetospheric boundary, $r\sim r_m$.
    Formation of a dense ring is
typical for $\Theta \lesssim 30^\circ$.
   At larger
$\Theta$, the ring does not form, and for $\Theta \gtrsim
60^\circ$, matter accretes to the star through equatorial  streams
(see Figures 16-18).

  Magnetic braking leads to a
significant departure of the angular velocity of the disk from
Keplerian in the region $r \lesssim 2-3$
   For $\Theta=60^\circ$ and
$\Theta=75^\circ$, the region of the disk out to $r\approx 2.5$
almost co-rotates with the star (see Figure 20c,d).

{\bf 5.}~ The inner regions of the disk are often  warped.
        This warp forms because
matter leaving the disk, start to
      move around magnetosphere,
thus forming a flow with axis close to that of $\rvecmu$.
   The region between the undisturbed disk
and the matter which co-rotates  with magnetosphere has a warped
shape.
   However, it has a different nature
compared to that predicted by theory of warped disks (see \S 4.4).

{\bf 6.}~  The angular momentum transport to the star was
investigated.
  For the considered small angular
velocity of the star, the angular momentum flux relative to the
$Z$ axis  gives a positive torque $N_z$ which acts  to spin-up of
the star for all $\Theta$.
  Angular
momentum is transported to the surface of the star by the magnetic
field  (as observed in our two-dimensional simulations, R02)
  The matter carries
about  1\% of the total angular momentum flux near the surface of
the star.

   The torques $N_x$ and $N_y$ were
calculated in the reference frame corotating with the star.
   They give information
about the long term evolution of the star's angular momentum $\bf
L$, and of $\bf\Omega$ and $\rvecmu$.
    For typical accretion rates
($\sim 10^{-7} - 10^{-9} M_\odot/$yr for a protostellar system),
the angular momentum of the star changes by a neglibible amount
during a rotating period of the inner region of the disk.
    This justifies our having
$\Theta$ fixed in the simulations.

\subsection{Observational Consequences}

     The described simulations are
important for understanding the physics of accreting magnetized
stars, for example, CTTSs, cataclysmic variables, and X-ray
pulsars.
   The simulations provide
a basis for understanding the photometric and spectral variability
the stars.
    Radiation in
different spectral bands may be associated with different regions,
the stellar surface, hot/cold spots on the stellar surface, and
the magnetospheric plasma.
    Knowledge of the
  magnetospheric flow is
essential to interpreting observed light curves, spectra, amd
temporal variations in the spectra.
   The presented simulations
may allow the estimation of $\Theta$ in different systems, and
may help to establish the main source of the radiation in
different wavebands.
    Some of predictions from this
work are:

{\bf 1.}~  Non-axisymmetric accretion and the formation of
accretion streams (which give hot spots on the star's surface) is
expected in almost all stars, because the limiting angle $\Theta$
at which non-axisymmetry appears is small ($\Theta \sim 2^\circ$).

{\bf 2.}~ At a fixed $\Theta$, the geometry of the magnetospheric
flow is different at different density levels.
    At a sufficiently low density,
all space is occupied by matter at this or larger densities.
     At a
larger density level,  ``windows'' appear in the accretion flow,
and at even larger densities, matter accretes along narrow
accretion streams.
    The variability of
different spectral lines and of different wavebands of the
continuum radiation is expected to depend on the density,
temperature and velocity distributions along the line of sight to
the star.
   Variation of the
accretion rate may also change the pattern of variability, because
at the same $\Theta$ the number of streams may temporary change.
Spiral structure may appear in the inner regions of the disk.

{\bf 3.}~ If the light curve has features with variability equal
to approximately ahalf of the main period, this may be a sign of
two rotating hot/cold spots, or, a sign of occultation of stellar
light by two magnetospheric streams. If the light curve suggest
occultations, then the inclination angle predicted by simulations
is relatively small, $\Theta < 30^\circ$. Relatively short periods
(few rotations) of the multi-stream features may appear, but most
of the time the two-stream features will dominate.
    Different observations
point to  such a possibility (e.g., Muzerolle, Hartmann \& Calvet
1998, Petrov et al. 2001).

{\bf 4.}  If the light curve  indicates precession of the
accretion streams, then the simulations point to the likelihood
that the inclination angle is  small, $\Theta \lesssim  10^\circ$.
    The variability pattern
for two precessing streams is characterized by an initial period
smaller than the half-period $P/2$, and a subsequent regular
increase of period to $P/2$.
     This
type of variability may be  rare because at small inclination
angles, the expected amplitude of variability is small.

{\bf 5.}  If the light  curve shows a quasi- period much shorter
than the period of rotation of the star (say, $\sim 1/4 P$), then
it will be a sign of  multiple streams which are typical for
intermediate inclination angles, $30^\circ <\Theta < 60^\circ$.
   Recently Muzerolle, Calvet, \&
Hartmann (2001), suggested that some observations of CTTSs may be
explained if the funnel flows have a multiple stream geometry.

{\bf 6.} At large inclination angles, $\Theta \gtrsim 60^\circ$,
matter also accretes in two streams, but they are located close to
the equatorial plane.
    In such a case
variability may be connected with occultation of stellar light by
the warped disk (as proposed by Bouvier et al. 1999, 2003). The
light curves are expected to be different from those where
occultation by magnetospheric streams dominate.

\acknowledgments This research was conducted using the resources
of the Cornell Theory Center, which receives funding  from Cornell
University, New York State, federal agencies, foundations, and
corporate partners. This work was supported in part by NASA grants
NAG5-9047, NAG5-9735, and by NSF grant AST-9986936. AVK and GVU
were partially supported by INTAS CALL2000-491 grant and by
Russian program ``Astronomy.'' J.V.W. was  supported by DOE
cooperative agreement DE-FC03-02NA00057.
  The authors thank Dr. Stinchcombe
for editing the manuscript.

\end{document}